\documentclass[]{aa}
\usepackage{graphicx}
\usepackage{natbib}
\usepackage{amssymb,amsmath}
\usepackage{rotating}
\usepackage{supertabular}

\newcommand{\CME}{_{\text{CME}}}
\newcommand{\p}{_{\text{p}}}
\newcommand{\s}{_{\text{s}}}
\newcommand{\eff}{_{\text{eff}}}
\newcommand{\J}{_{\text{J}}}
\newcommand{\cc}{_{\text{c}}}
\newcommand{\ssun}{_\odot}
\newcommand{\sstar}{_{\star}}
\newcommand{\tis}{\tau_{\text{sync}}}
\newcommand{\kep}{_{\text{orbit}}}

\begin{document}

\title{Predicting low-frequency radio fluxes of known extrasolar planets\thanks{Table 1 is only available in electronic 
form at the CDS via anonymous ftp to cdsarc.u-strasbg.fr (130.79.128.5) or via http://cdsweb.u-strasbg.fr/cgi-bin/qcat?J/A+A/}
}

  \author{J.--M. Grie{\ss}meier
          \inst{1}
          \and
          P. Zarka
          \inst{1}
          \and
          H. Spreeuw\inst{2}
 }


   \institute{
   	LESIA, Observatoire de Paris, CNRS, UPMC, Universit\'{e} Paris Diderot; 5 Place Jules Janssen, 92190 Meudon, France\\
              \email{jean-mathias.griessmeier@obspm.fr, philippe.zarka@obspm.fr}
    \and
            Astronomical Institute ``Anton Pannekoek'', Kruislaan 403, 1098 SJ  Amsterdam, Netherlands\\
            \email{hspreeuw@science.uva.nl}
}

\date{Version of \today}

\abstract{
Close-in giant extrasolar planets (``Hot Jupiters'') are believed to be strong emitters in the decametric radio range.
}
{We present the expected characteristics of the low-frequency magnetospheric 
radio emission of all currently known extrasolar planets, including the  maximum emission frequency and the expected radio flux. We also discuss the escape of exoplanetary radio emission from 
the vicinity of its source,
which imposes additional constraints on detectability.  
}
{We compare the different predictions obtained with all four existing analytical models for all currently known exoplanets. We also take care to use realistic values for all input parameters.}
{
The four different models for planetary radio emission 
lead to very different results. The largest fluxes are found for the \textit{magnetic} energy model, followed by the \textit{CME} model and the \textit{kinetic} energy model (for which our results are found to be much less optimistic than those of previous studies). The \textit{unipolar interaction} model does not predict any observable emission for the present exoplanet census. 
We also give estimates for the planetary magnetic dipole moment of all currently known extrasolar planets, which will be useful for other studies.
}
{Our results show that observations of exoplanetary radio emission are feasible, but that the number of promising targets is not very high.
The catalog of targets will be particularly useful for current and future radio observation campaigns
(e.g.~with the VLA, GMRT, UTR-2 and with LOFAR).}

\maketitle

\section{Introduction}
In the solar system, all strongly magnetised planets are known to be intense nonthermal radio 
emitters. For a certain class of extrasolar planets (the so-called Hot Jupiters), an analogous, 
but much more intense radio emission is expected.
In the recent past, such exoplanetary radio emission has become an active field of research, 
with both theoretical studies and ongoing observation campaigns.

Recent theoretical studies have shown that a large variety of effects have to be considered, e.g.~kinetic, magnetic and 
unipolar interaction between the star (or the stellar wind) and the planet, the influence of the 
stellar age, the potential role of stellar CMEs, and the influence of different stellar wind 
models. 
So far, there is no single publication in which all of these aspects are put together and 
where the different interaction models are compared extensively.
We also discuss the escape of exoplanetary radio emission from its planetary system,
which depends on the local stellar wind parameters. As will be shown, this is an additional constraint for detectability, making the emission from several planets impossible to observe. 

The first observation attempts go back at least to \citet{Yantis77}.
At the beginning, such observations were necessarily unguided ones, as exoplanets had not yet been discovered. 
Later observation campaigns concentrated on known exoplanetary systems. So far, no detection has
been achieved. A list and a comparison of past observation attempts can be found elsewhere 
\citep[][]{Griessmeier51PEG05}. 
Concerning ongoing and future observations, studies are performed or planned at the VLA \citep{Lazio04}, 
GMRT \citep{Majid05,Winterhalter06}, UTR2 \citep{Ryabov04}, and at LOFAR \citep{Farrell04}. 
To support these observations and increase their efficiency, it is important to identify the most
promising targets. 

The target selection for radio observations is based on theoretical estimates which aim at the prediction of the main characteristics of the exoplanetary radio emission. The two most important characteristics are the 
maximum frequency of the emission and the expected radio flux. The first predictive studies 
\citep[e.g.][]{Zarka97,Farrell99}
concentrated on only a few exoplanets. A first catalog containing estimations for radio emission of a large number of exoplanets was presented by \citet{Lazio04}. This catalog included 118 planets (i.e.~those known as of 2003, July 1) and 
considered radio emission energised by the kinetic energy of the stellar wind (i.e.~the \textit{kinetic model}, see below). Here, we present 
a much larger list of targets (i.e.~197 exoplanets found by radial velocity and/or transit searches as of 2007, January 13, taken from http://exoplanet.eu/), and compare
the results obtained by all four currently existing interaction models, not all of which were known at the time of
the previous overview study. As a byproduct of the radio flux calculation, we obtain estimates for the planetary magnetic dipole moment of all currently known extrasolar planets. These values will be useful for other studies as, e.g., star-planet interaction or atmospheric shielding.

To demonstrate which stellar and planetary parameters are required for the estimation of exoplanetary
radio emission, some theoretical results are briefly reviewed (section \ref{sec:theory}). 
Then, the sources for the different parameters (and their default values for the case where no measurements 
are available) are presented (section \ref{sec:modelling}). 
In section \ref{sec:results}, we present our estimations for exoplanetary radio emission. This section also includes estimates for planetary magnetic dipole moments.
Section \ref{sec:conclusion} closes with a few concluding remarks.

\section{Exoplanetary radio emission theory} \label{sec:theory}

\subsection{Expected radio flux} \label{sec:flux}

In principle, there are four different types of interaction between a planetary obstacle and the ambient stellar wind, 
as both the stellar wind and the planet can either be magnetised or unmagnetised. 
\citet[][Table 1]{Zarka06PSS} show that for three of these four 
possible situations 
intense nonthermal radio emission is possible. Only in the case of an 
unmagnetised stellar wind interacting with an unmagnetised body no intense radio emission is possible.

In those cases where strong emission is possible, the expected radio flux depends on the source
of available energy. In the last years, four different energy sources were suggested:
a) In the first model, the input power $P_{\text{input}}$  into the magnetosphere is assumed to be  
proportional to the total \textit{kinetic energy} flux of the solar wind protons impacting on 
the magnetopause
\citep[][]{Desch84,Zarka97,Farrell99,Zarka01,Farrell04,Lazio04,Stevens05,Griessmeier05,GriessmeierPREVI,GriessmeierPSS06} 
b) Similarly, the input power $P_{\text{input}}$  into the magnetosphere can be assumed to be proportional 
to the \textit{magnetic energy} flux or electromagnetic Poynting flux of the interplanetary magnetic field 
\citep{Zarka01,Farrell04,Zarka04a,Zarka06PREVI,Zarka06PSS}.
From the data obtained in the solar system, it is not possible to distinguish which of these 
models is more appropriate \citep[the constants of proportionality implied in the relations given below are not well known, see][]{Zarka01}, 
so that both models have to be considered. 
c) For unmagnetised or weakly magnetised planets, one may apply the \textit{unipolar 
interaction} model. In this model, the star-planet system can be seen as a giant analog to the 
Jupiter-Io system \citep{Zarka01, Zarka04a,Zarka06PREVI,Zarka06PSS}. Technically, this model is
very similar to the magnetic energy model, but the source location is very different: Whereas in
the kinetic and in the magnetic model, the emission is generated near the planet, in the 
unipolar interaction case a large-scale current system is generated and the radio emission is 
generated in the stellar wind between the star and the planet. Thus, the emission can originate 
from a location close to the stellar surface, close to the planetary surface, or at any point between the two. This is possible in those cases where the solar wind speed is lower than the Alfv\'{e}n velocity 
\citep[i.e.~for close-in planets, see e.g.][]{Preusse05}. Previous studied have indicated that this emission is unlikely to be detectable,
except for stars with an extremely strong magnetic field \citep{Zarka01,Zarka04,Zarka06PREVI,Zarka06PSS}. 
Nevertheless, we will check whether this type of emission is possible for the known exoplanets.
d) The fourth possible energy source is based on the fact that 
close-in exoplanets are expected to be subject to frequent and violent stellar eruptions  \citep{Khodachenko05} similar to solar coronal mass ejections (CMEs). As a variant to the kinetic energy model, the \textit{CME} model assumes that the energy for 
the most intense planetary radio emission is provided by CMEs. During periods of such CME-driven 
radio activity, considerably higher radio flux levels can be achieved than during quiet 
stellar conditions \citep{GriessmeierPREVI,GriessmeierPSS06}. For this reason, this model is 
treated separately.

For the \textit{kinetic energy} case, the input power was first derived by 
\citet[][]{Desch84}, who found that it is given by 
\begin{equation}
	P_{\text{input,kin}} \propto  n v\eff^3 R\s^2. \label{eq:Pin:kin}
\end{equation} 
In eq.~(\ref{eq:Pin:kin}), $n$ is the stellar wind density at the planetary orbit, $v\eff$ is the velocity of the stellar wind in the reference frame of the planet (i.e.~including the aberration due to the orbital velocity of the planet, which is not negligible for close-in planets), and $R\s$ denotes the magnetospheric standoff distance. 

The \textit{magnetic energy} case was first discussed by \citet{Zarka01}.
Here, the input power is given by 
\begin{equation}
	P_{\text{input,mag}} \propto  v\eff B_{\perp}^2 R\s^2 \label{eq:Pin:mag}
\end{equation}
In eq.~(\ref{eq:Pin:mag}), $v\eff$ is the velocity of the stellar wind in the reference frame of the planet, 
$B_{\perp}$ if the component of the interplanetary magnetic field (IMF) perpendicular to the stellar wind flow  in the reference frame of the planet, and $R\s$ denotes the magnetospheric standoff distance.

For the \textit{unipolar interaction} case \citep{Zarka01}, the input power is given by 
\begin{equation}
	P_{\text{input,unipolar}} \propto  v\eff B_{\perp}^2 R_{\text{ion}}^2 \label{eq:Pin:uni}
\end{equation}
Eq.~(\ref{eq:Pin:uni}) is identical to eq.~(\ref{eq:Pin:mag}), except that the obstacle is not the planetary magnetosphere, but its ionosphere, so that $R\s$ is replaced by $R_{\text{ion}}$, the radius of the planetary ionosphere.

\textit{CME}-driven radio emission was first calculated by \citet{GriessmeierPREVI}. In that case, the input power is given by 
\begin{equation}
	P_{\text{input,kin,CME}} \propto  n\CME v_{\text{eff,CME}}^3 R\s^2. \label{eq:Pin:kin:CME}
\end{equation}
Eq.~(\ref{eq:Pin:kin:CME}) is identical to eq.~(\ref{eq:Pin:kin}), except that the stellar wind density and velocity are replaced by the corresponding values encountered by the planet during a CME.

A certain fraction $\epsilon$ of the input power $P_{\text{input}}$ given by eq.~(\ref{eq:Pin:kin}), (\ref{eq:Pin:mag}), (\ref{eq:Pin:uni}) or (\ref{eq:Pin:kin:CME})  is thought to be dissipated within the magnetosphere:
\begin{equation}
	P_d=\epsilon P_{\text{input}}
\end{equation} 

Observational evidence suggests that the amount of power emitted by radio waves 
$P_{\text{rad}}$ is roughly proportional to the power input $P_{\text{input}}$ 
\citep[see, e.g.][Figure 6]{Zarka06PSS}. This can be written as:
\begin{equation}
	P_{\text{radio}}=\eta_{\text{radio}} P_d=\eta_{\text{radio}}\epsilon P_{\text{input}}
\end{equation}
As $P_d$ cannot be measured directly, one correlates the observed values of $P_{\text{radio}}$ with the calculated (model dependent) values of $P_{\text{input}}$. 
Thus, one replaces $P_{\text{input}}$ by $P_{\text{radio}}$ on the left-hand side of the proportionalities given by~(\ref{eq:Pin:kin}), (\ref{eq:Pin:mag}), (\ref{eq:Pin:uni}) and (\ref{eq:Pin:kin:CME}). The proportionality constant is determined by comparison with Jupiter.
The analysis of the jovian radio emission allows to define three values for the typical 
radio spectrum:  (a) the power during \textit{average conditions}, (b) the average power during 
periods of \textit{high activity}, and (c) the  \textit{peak power} \citep{Zarka04}. In this work, we will use the average power during periods of high activity as a reference value for all four cases, with 
$P_{\text{radio,J}}=2.1 \cdot 10^{11}$~W.

The radio flux  $\Phi$ seen by an observer at a distance $s$ from the emitter is related to the emitted radio power $P_{\text{radio}}$ by \citep{GriessmeierPSS06}:
\begin{equation}
  \Phi=\frac{P_{\text{radio}}}
  {\Omega s^2 \Delta f}
	=\frac{4 \pi^2 m_e R\p^3 P_{\text{radio}}} {e \mu_0 \Omega s^2 \mathcal{M}}.  \label{eq:Phi:s}
\end{equation}
Here, $\Omega$ is the solid angle of the beam of the emitted radiation 
\citep[$\Omega= 1.6$ sr, see][]{Zarka04}, and $\Delta f$ is the bandwidth of the emission. We use $\Delta f=f\cc^\text{max}$ 
\citep{GriessmeierPSS06}, where $f\cc^\text{max}$ is the maximum cyclotron frequency.
Depending on the model, $P_{\text{radio}}$ is given by eq.~(\ref{eq:Pin:kin}), (\ref{eq:Pin:mag}), (\ref{eq:Pin:uni}) or (\ref{eq:Pin:kin:CME}). 
The maximum cyclotron frequency 
$f\cc^\text{max}$ is determined by the maximum magnetic field strength $B\p^{\text{max}}$ close 
to the polar cloud tops \citep{Farrell99}:
\begin{equation}
	f\cc^\text{max}= \frac{eB\p^\text{max}}{2\pi m_e}=\frac{e \mu_0 \mathcal{M}}{4\pi^2 m_e R\p^3}
	\approx 24 \,\text{MHz} \, \frac{\widetilde{\mathcal{M}}}{\widetilde{R\p}^3}
	.\label{eq:f}
\end{equation}
Here, $m_e$ and $e$ are the electron mass and charge, $R\p$ is the planetary radius, $\mu_0$ is the magnetic permeability of the vacuum, and $\mathcal{M}$ is the planetary magnetic dipole moment. 
$\widetilde{\mathcal{M}}$ and $\widetilde{R\p}$ denote the planetary magnetic moment and its radius relative to the respective value for Jupiter, e.g.~$\widetilde{\mathcal{M}}=\mathcal{M}/\mathcal{M}\J$, with $\mathcal{M}\J=1.56 \cdot 10^{27}$ Am$^2$ \citep{Cain95} and $R\J=71492$ km.

The radio flux expected for the four different models according to eqs.~(\ref{eq:Pin:kin}) to (\ref{eq:Phi:s}) and the 
maximum emission frequency according to (\ref{eq:f}) are 
calculated in section \ref{sec:results} for all known exoplanets.

\subsection{Escape of radio emission} \label{sec:escape}

To allow an observation of exoplanetary radio emission, it is not sufficient to have a high enough emission power at the source and emission in an observable frequency range. As an additional requirement, it has to be checked that the emission can propagate from the source to the observer. This is not the case if the emission is absorbed or trapped (e.g.~in the stellar wind 
in the vicinity of the radio-source),
which happens whenever the plasma frequency 
\begin{equation}
	f_{\text{plasma}}=\frac{1}{2\pi}\sqrt{\frac{ne^2}{\epsilon_0 m_e}}  \label{eq:fplasma}
\end{equation}	
is higher than the emission frequency at any point between the source and the observer. Thus, the condition of observability is
\begin{equation}
	f_{\text{plasma}}^\text{max}<f\cc^\text{max} \label{eq:escape},
\end{equation}
where $f\cc^\text{max} $ is taken at the radio source (e.g.~the planet), whereas 
$f_{\text{plasma}}^\text{max}$ is evaluated along the line of sight.
As the density of the electrons in the stellar wind $n$ decreases with the distance to the star,  
this condition is more restrictive at the orbital distance of the planet than further out.
Thus, it is sufficient to check whether condition (\ref{eq:escape}) is satisfied at the location of the 
radio-source
(i.e.~for $n=n(d)$, where $d$ is the distance from the star to the radio-source). 
In that case, the emission can escape from the planetary system and reach distant observers. 
In section \ref{sec:results}, condition (\ref{eq:escape}) is checked for all known exoplanets at their orbital distance.

Note however that, depending on the line of sight, not all observers will be able to see the planetary emission at
all times. For example, the observation of a secondary transit implies that the line of sight passes very close to the planetary host star, where the plasma density is much higher. For this reason, some parts of the orbit may be unobservable even for planets for which eq.~(\ref{eq:escape}) is satisfied.

\subsection{Radiation emission in the unipolar interaction model} \label{sec:unipolar}

An additional constraint arises because certain conditions are necessary for the generation of radio emission.
Planetary radio emission is caused by the cyclotron maser instability (CMI). This mechanism is only efficient
in regions where the ratio between the electron plasma frequency and the electron cyclotron frequency is small enough. This condition can be written as 
\begin{equation}
	\frac{f_{\text{plasma}}}{f\cc} \lesssim 0.4 \label{eq:f:generation},
\end{equation}
where the electron cyclotron frequency $f\cc$ 
is defined by the local magnetic field
\begin{equation}
	{f\cc} = \frac{eB}{2\pi m_e}
\end{equation}
and $f_{\text{plasma}}$ is given by  
eq.~(\ref{eq:fplasma}).
Observations seem to favor a critical frequency ratio close to the $0.1$,  
while theoretical work supports a critical frequency ratio close to $0.4$ 
\citep{LeQueau85,Hilgers92,Zarka01cutoff}.
Fundamental O mode or second harmonic O and X mode emission are possible also for larger 
frequency ratios, but are much less efficient \citep{Treumann00,Zarka06PSS}. To avoid ruling out potential emission, we use the largest possible frequency ratio, i.e.~$\frac{f_{\text{plasma}}}{f\cc} \le 0.4$.

The condition imposed by eq.~(\ref{eq:f:generation}) has to be fulfilled for any of the 
four models presented in section \ref{sec:flux}.
For the three models where the radio emission is generated directly in the planetary magnetosphere,
$n$ decreases much faster with distance to the planetary surface than $B$, so that eq.~(\ref{eq:f:generation}) can always be fulfilled.
For the \textit{unipolar interaction} model, the emission takes place in the stellar wind, and the electron density $n$ 
can be obtained from the model of the stellar wind.
In this case, it is not a priori clear where radio emission is possible. It could be generated anywhere between the star and the planet. 
In section \ref{sec:results}, we will check separately for each planetary system whether unipolar interaction satisfying 
eq.~(\ref{eq:f:generation}) is possible at any location
between the stellar surface and the planetary orbit. 
\section{Required parameters} \label{sec:modelling}

In the previous section, it has been shown that the detectability of planetary radio emission
depends on a few planetary parameters:
\begin{itemize}
	\item	the planetary radius $R\p$
	\item	the planetary magnetic moment $\mathcal{M}$
	\item	the size of the planetary magnetosphere $R\s$ 
	\item	the size of the planetary ionosphere $R_{\text{ion}}$
	\item	the stellar wind density $n$ and its velocity $v\eff$
	\item	the stellar magnetic field (IMF) $B_\perp$ perpendicular to the stellar wind flow in the frame of the planet
	\item	the distance of the stellar system (to an earth-based observer) $s$
	\item the solid angle of the beam of the emitted radiation $\Omega$
\end{itemize}

The models used to infer the missing stellar and planetary quantities require the knowledge of a few 
additional planetary parameters. These are the following:
\begin{itemize}
	\item the planetary mass $M\p$
	\item	the planetary radius $R\p$
	\item its orbital distance $d$
	\item	the planetary rotation rate $\omega$
	\item	the stellar magnetic field (IMF) components $B_r, B_\phi$
	\item the stellar mass $M_\star$
	\item the stellar radius $R_\star$
	\item	the stellar age $t_\star$
\end{itemize}

In this section, we briefly describe how these each of these quantities can be obtained. 

\subsection{Basic planetary parameters}

As a first step, basic planetary characteristics have to be evaluated:
 \begin{itemize}
	\item   $d, \omega_{\text{orbit}}$ and $s$ are directly taken from the Extrasolar 
		Planets Encyclopaedia at {\tt http://exoplanet.eu}, as well as the observed 
		mass $M_{\text{obs}}$ and the orbital eccentricity $e$.
		Note that in most cases the observed mass is the ``projected mass'' of the planet,
		i.e.~$M_{\text{obs}}= M\p \sin i$, where $i$ is the angle of inclination of the 
		planetary orbit with respect to the observer.
	\item	For eccentric planets, we calculate the periastron from the semi-major axis $d$ 
		and the orbital eccentricity $e$: $d_{\text{min}}=d/(1-e)$. Thus, the results for 
		planetary radio emission apply to the periastron.
	\item	For most exoplanets, the planetary mass is not precisely known. Instead, usually 
		only the projected mass $M\p \sin i$ is accessible to measurements, where $i$ is 
		the inclination of the planetary orbit with respect to the observer. This 
		projected mass is taken from {\tt http://exoplanet.eu} and converted to the median 
		value of the mass: 
		$\text{median}(M\p)=M_{\text{obs}} \cdot \text{median}(\frac{1}{\sin i})
		= \sqrt{4/3} \cdot M_{\text{obs}}
		\approx 1.15 \, M_{\text{obs}}$.
	\item	For transiting planets, $R\p$ is taken from original publications.
		For non-transiting planets, the planetary radius is $R\p$ not known. 
		In this case, we estimate the planetary radius based on its mass $M\p$ and 
		orbital distance $d$, as explained in appendix \ref{sec:appendix}. 
		The radius of a ``cold'' planet of mass $M\p$ is given by
		\begin{equation}
			R\p(d=\infty)= 
			\frac{ \left( \alpha M\p \right)^{1/3}}{1+\left( \frac{M\p}{M_{\text{max}}} \right)^{2/3}}
			\approx 1.47 R\J \,\frac{  \widetilde{M\p} ^{1/3}}{1+\left( \frac{\widetilde{M\p}}{\widetilde{M_{\text{max}}}} \right)^{2/3}}
		\end{equation}
		with $\alpha=6.1 \cdot 10^{-4}$ m$^3$ kg$^{-1}$ (for a planet with the same composition as 
		Jupiter) and $M_{\text{max}}=3.16 \, M\J$. 
		Again, $\widetilde{M\p}$ and $\widetilde{M_{\text{max}}}$ denote values relative to the 
		respective value for
		Jupiter (using ${M}\J=1.9 \cdot 10^{27}$ kg).
		The radius of an irradiated planet is then given by 
		\begin{equation}
			\frac{R\p(d)}{R\p(d=\infty)}=\frac{\widetilde{R\p}(d)}{\widetilde{R\p}(d=\infty)}=
			 \cdot \left[ 1+0.05\left(\frac{T_{\text{eq}}}{T_0}\right)^{\gamma} \right] 
			\label{eq:temp}
		\end{equation}	
		where $T_{\text{eq}}$ is the equilibrium temperature of the planetary surface. The coefficients 
		$T_0$ and $\gamma$ depend on the planetary mass (see appendix \ref{sec:appendix}).
\end{itemize}	
 
\subsection{Stellar wind model} \label{sec:stellarwind}

The stellar wind density $n$ and velocity $v\eff$ encountered by a planet are key 
parameters defining the size of the magnetosphere and thus the energy flux available to  
create planetary radio emission. As these stellar wind parameters strongly depend on the stellar age, 
the expected radio flux is a function of the estimated age of the exoplanetary host star 
\citep{Stevens05,Griessmeier05}. 
At the same time it is known that at close distances the stellar wind velocity has not yet 
reached the value it has at larger orbital distances. For this reason, a distance-dependent 
stellar wind models has to be used to avoid overestimating the expected planetary radio flux 
\citep{GriessmeierPHD06,GriessmeierPSS06}. 

It was shown \citep{GriessmeierPSS06} that for stellar ages $>0.7$ Gyr, the
radial dependence of the stellar wind properties can be described by the
stellar wind model of \citet{Parker58}, and that the more complex model of \citet{Weber67} is not required.
In the Parker model, the interplay between stellar gravitation and pressure gradients leads to 
a supersonic gas flow for sufficiently large substellar distances $d$. 
The free parameters are the coronal temperature
and the stellar mass loss. They are indirectly chosen by setting the stellar wind conditions at 1 AU.
More details on the model 
can be found elsewhere \citep[e.g.][]{Mann99,PreussePHD05,GriessmeierPHD06}. 

The dependence of the stellar wind density $n$ and velocity and $v\eff$ on the age of the stellar system is based on observations of astrospheric absorption features of stars with different ages. 
In the region between the astropause and the astrospheric bow shock (analogs to the heliopause
and the heliospheric bow shock of the solar system), the partially ionized local 
interstellar medium (LISM) is heated and compressed. Through charge exchange processes,
a population of neutral hydrogen atoms with high temperature is created. The characteristic 
Ly$\alpha$ absorption (at 1216 $\text{\AA}$) of this population was detectable with the high-resolution 
observations obtained by the Hubble Space Telescope (HST).
The amount of absorption depends on the size of the astrosphere, which is a function of the 
stellar wind characteristics. 
Comparing the measured absorption to that calculated by hydrodynamic codes, these measurements 
allowed the first empirical estimation of the evolution of the stellar mass loss rate as a 
function of stellar age \citep{Wood02,Wood04,Wood05}. It should be noted, however, that the resulting estimates are only valid for stellar ages $\ge0.7$ Gyr \citep{Wood05}.
From these observations, \citep{Wood05} calculate the age-dependent density of the stellar wind under the assumption of an age-independent stellar wind velocity.
This leads to strongly overestimated stellar wind densities, especially for young stars \citep{Griessmeier05,Holzwarth07}. For this reason, we combine these results with the model for the 
age-dependence of the stellar wind velocity of  \citet{Newkirk80}. One obtains 
\citep{GriessmeierPSS06}:
\begin{equation}
  v(1 \text{AU}, t)= v_0 \left( 1+\frac{t}{\tau}\right)^{-0.43}.
  \label{eq:scaleV}
\end{equation}
The particle density can be determined to be
\begin{equation}
  n(1 \text{AU}, t)= n_0 \left( 1+\frac{t}{\tau}\right)^{-1.86\pm0.6}.
  \label{eq:scaleN}
\end{equation}
with 
$v_0=3971$ km/s, $n_0=1.04\cdot10^{11}\text{ m}^{-3}$ and
$\tau=2.56\cdot10^7 \text{yr}$. 

For planets at small orbital distances, the keplerian velocity of the planet moving around its
star becomes comparable to the radial stellar wind velocity. Thus, 
the interaction of the stellar wind with the planetary magnetosphere should be calculated using
the effective velocity of the stellar wind plasma relative to the planet, which takes into 
account this ``aberration effect''
\citep{Zarka01}.
For the small orbital distances relevant for Hot Jupiters, the planetary orbits are circular
because of tidal dissipation \citep{Goldreich66,DobbsDixon04,Halbwachs05}.
For circular orbits, the orbital velocity $v\kep$ is perpendicular to 
the stellar wind velocity $v$, and its value is given by Kepler's law.
In the reference frame of the planet, the stellar wind velocity then is given by
\begin{equation}
	v\eff=\sqrt{v^2_{\text{orbit}}+v^2}. \label{eq:veff}
\end{equation}

Finally, for the magnetic energy case, the interplanetary magnetic field ($ B_r, B_\phi$) is 
required.
At 1 AU, the average field strength of the interplanetary magnetic field is 
$B_{\text{imf}}\approx 3.5\,\text{nT}$ \citep{Mariani90,Proelss04}.
According to the Parker stellar wind model \citep{Parker58}, 
the radial component of the interplanetary magnetic field decreases as 
\begin{equation}
	B_{\text{imf},r}(d)=B_{r,0} \left(\frac{d}{d_0}\right)^{-2}. \label{eq:Bimfr}
\end{equation}	
This was later confirmed by Helios measurements. 
One finds $B_{r,0}\approx 2.6$ nT and $d_0=1$ AU \citep{Mariani90,Proelss04}. 
At the same time, the azimuthal component 
$B_{\text{imf},\varphi}$ behaves as
\begin{equation}
	B_{\text{imf},\varphi}(d)=B_{\varphi,0} \left(\frac{d}{d_0}\right)^{-1}, 
	\label{eq:Bimfphi}
\end{equation}
with $B_{\varphi,0}\approx2.4$ nT \citep{Mariani90,Proelss04}.
The average value of $B_{\text{imf},\theta}$ vanishes ($B_{\text{imf},\theta}\approx 0$).
From $B_{\text{imf},r}(d)$ and $B_{\text{imf},\varphi}(d)$, the stellar magnetic field (IMF) $B_\perp(d)$ perpendicular to the stellar wind flow in the frame of the planet can be calculated \citep{Zarka06PSS}:
\begin{equation}
	B_\perp = \sqrt{B_{\text{imf},r}^2+B_{\text{imf},\varphi}^2} \left| \sin\left(\alpha-\beta \right) \right| 
\end{equation}
with
\begin{equation}
	\alpha = \arctan \left( \frac{B_{\text{imf},\varphi}}{B_{\text{imf},r}}  \right) 
\end{equation}
and
\begin{equation}
	\beta = \arctan \left( \frac{v_{\text{orbit}}}{v}  \right).
\end{equation}

We obtain the stellar magnetic field $B\sstar$ relative to the solar magnetic field $B\ssun$ under the assumption that it is inversely proportional 
to the rotation period $P\sstar$ \citep{CollierCameron94,GriessmeierPSS06}:   
\begin{equation}
	\frac{B\sstar}{B\ssun} = \frac{P\ssun}{P\sstar},
\end{equation}
where we use $P\ssun=25.5$~d and take $B_\odot=1.435\cdot 10^{-4}$ T as the reference 
magnetic field strength at the solar surface \citep{Preusse05}.
The stellar rotation period $P\sstar$ is calculated from the stellar age $t$ \citep{Newkirk80}:
\begin{equation}
	P\sstar\propto \left(1+\frac{t}{\tau}\right)^{0.7},
\end{equation} 
where the time constant $\tau$ is given by $\tau=2.56\cdot10^7$ yr 
\citep[calculated from][]{Newkirk80}. 

Note that first measurements of stellar magnetic fields for planet-hosting stars are just becoming
available from the spectropolarimeter ESPaDOnS \citep{Catala07}. This will lead to an improved understanding of stellar magnetic 
fields, making more accurate models possible in the future.

\subsection{Stellar CME model}

For the CME-driven radio emission, the stellar wind parameters $n$ and $v\eff$ are effectively replaced by the corresponding CME parameters $n\CME$ and $v_\text{eff,CME}$, potentially leading to much more intense  radio emission than those driven by the kinetic energy of the steady stellar wind \citep{GriessmeierPHD06,GriessmeierPREVI,GriessmeierPSS06}.

These CME parameters are estimated by \citet{Khodachenko05}, who combine in-situ measurements near the sun 
(e.g.~by Helios) with remote solar observation by SoHO. 
Two interpolated limiting cases 
are given, denoted as {\it weak} and {\it strong} CMEs, respectively. These two classes have a 
different dependence of the average density on the distance to the star $d$. 
In the following, these quantities will be labeled $n\CME^w(d)$ and $n\CME^s(d)$, respectively.

For weak CMEs, the density $n\CME^w(d)$ behaves as
\begin{equation}
  n\CME^w(d) = n_{\text{CME},0}^w \left( d/d_0 \right)^{-2.3} \label{eq:sme:nw}
\end{equation}  
where the density at $d_0=1$ AU is given by 
$n_{\text{CME},0}^w=n\CME^w(d=d_0)=4.9 \cdot 10^6$ m$^{-3}$.

For strong CMEs, \citet{Khodachenko05} find
\begin{equation}
  n\CME^s(d) = n_{\text{CME},0}^s \left(  d/d_0 \right)^{-3.0} \label{eq:sme:ns}
\end{equation}  
with $n_{\text{CME},0}^s=n\CME^s(d=d_0)=7.1 \cdot 10^6$ m$^{-3}$, and $d_0=1$ AU.

As far as the CME velocity is concerned, one has to note that individual CMEs have very 
different velocities. However, the {\it average} CME velocity $v$ is approximately independent 
of the subsolar distance, and is similar for both types of CMEs:
\begin{equation}
  v\CME^w=v\CME^s=v\CME \approx 500 \, \text{km/s}. \label{eq:sme:v} 
\end{equation}
Similarly to the steady stellar wind the CME velocity given by eq.~(\ref{eq:sme:v}) 
has to be corrected for the orbital motion of the planet:
\begin{equation}
  v_{\text{eff,CME}}=\sqrt{\frac{M_\star G}{d}+v\CME^2}. \label{eq:veff:CME}
\end{equation}

In addition to the density and the velocity, the temperature of the plasma in a coronal mass
ejection is required for the calculation of the size of the magnetosphere.
According to \citet{Khodachenko05,Khodachenko06PSS}, the front region of a CME 
consists of hot, coronal material 
($T \approx 2$ MK). This region may either be followed by relatively cool prominence material 
($T \approx 8000$ K), or by hot flare material ($T \approx 10$ MK). In the following, the
temperature of the leading region of the CME will be used, i.e.~$T\CME=2$ MK. 

\subsection{Planetary magnetic moment and magnetosphere} \label{sec:protation}

For each planet, the value of the planetary magnetic moment $\mathcal{M}$ is estimated by taking the 
geometrical mean of the maximum and minimum result obtained by different scaling laws. The associated uncertainty was discussed by \citet{GriessmeierPSS06}.
The different scaling laws are compared, e.g., by 
\citet[][]{Farrell99}\footnote{containing a typo in the sixth equation of the appendix.},
\citet{Griessmeier04} and 
\citet{GriessmeierPHD06}. In order
to be able to apply these scaling laws, some assumptions on the planetary size and structure are
required. The variables required in the scaling laws are $r\cc$ (the radius of the dynamo region
within the planet), $\rho\cc$ (the density within this region), $\sigma$ (the conductivity within
this region) and $\omega$ (the planetary rotation rate).

\paragraph{The size of the planetary core $r\cc$ and its density $\rho\cc$}

The density
profile within the planet $\rho(r)$ 
is obtained by describing the planet as a polytropic gas sphere, using the solution of the 
Lane-Emden equation \citep{Chandrasekhar57,Sanchez04}:
\begin{equation}
	\rho(r)=\left( \frac{\pi M\p}{4 R\p^3} \right) \frac{\sin\left(\pi \frac{r}{R\p}\right)}{\left(\pi \frac{r}{R\p}\right)}.
\end{equation}
The size of the planetary core $r\cc$ is found by searching for the value of $r$ where the density 
$\rho(r)$ becomes large 
enough for the transition to the liquid-metallic phase \citep{Sanchez04,Griessmeier05,GriessmeierPHD06}.
The transition was assumed to occur at a density of 700 kg/m$^3$, which is consistent with the 
range of parameters given by \citet{Sanchez04}. 
For Jupiter, we obtain $r\cc=0.85\, R\J$.

The average density in the dynamo region 
$\rho\cc$ is then obtained by averaging the density $\rho(r)$ over the range $0\le r\le r\cc$. 
For Jupiter, we obtain $r\cc\approx 1800$ kg m$^{-3}$.
\paragraph{The planetary rotation rate $\omega$} 

Depending on the orbital distance of the planet and the timescale for synchronous
rotation $\tis$  (which is derived in appendix \ref{sec:tlocking}),
three cases can be distinguished:
\begin{enumerate}
	\item	For planets at small enough distances for which the timescale for tidal locking 
		is small (i.e.~$\tis \le 100$ Myr), the rotation period is taken to be 
		synchronised with the orbital period 
		($\omega=\omega_\text{f}\approx\omega_{\text{orbit}}$), which is known 
		from measurements. This case will be denoted by ``TL''.
		Typically, this results in smaller rotation rates for tidally locked planets than for freely
		rotating planets.
	\item	Planets with distances resulting in
		$100 \ \text{Myr} \le \tis \le 10\ \text{Gyr}$
		may or may not be subject to tidal locking. This will, for
		example, depend on the exact age of the planetary system, which is typically in
		the order of a few Gyr. For this reason, 
		we calculate the expected characteristics of these ``potentially locked'' planets 
		\textit{twice}: once with tidal locking 
		(denoted by ``(TL)'') and once without tidal locking (denoted by ``(FR)'').
	\item	For planets far away from the central star, the timescale for tidal locking 
		is very large. For planets with $\tis \ge 10$ Gyr, the
		effect of tidal interaction can be neglected. In this case, the planetary rotation
		rate can be assumed to be equal to the initial rotation rate $\omega_\text{i}$,
		which is assumed to be equal to the current rotation rate of Jupiter,
		i.e.~$\omega=\omega\J$
		with $\omega\J=1.77\cdot10^{-4}$ s$^{-1}$.
		This case will be denoted by ``FR''.		
\end{enumerate}

Note that tidal interaction does not perfectly synchronise the planetary 
rotation to its orbit. Thermal atmospheric tides resulting from 
stellar heating can drive planets away from synchronous rotation 
\citep{Showman02,Correia03,Laskar04}.
According to \citet{Showman02}, the corresponding 
error for $\omega$ could be as large as a factor of two. 
On the basis of the example of $\tau$ Bootis b, \citet{GriessmeierPSS06} show that the effect of imperfect tidal locking (in combination with the spread of the results found by different scaling laws) can lead to magnetic moments and thus emission frequencies up to a factor 2.5 higher than for the nominal case. Keeping this ``error bar'' in mind, we will 
nevertheless consider only the reference case in this work.
\paragraph{The conductivity in the planetary core $\sigma\cc$}

Finally, the conductivity in the dynamo region of extrasolar planets remains to be evaluated. 
According to \citet{Nellis00}, the electrical conductivity remains constant throughout
the metallic region. For this reason, it is not necessary to average over the volume of the
conducting region.
As the magnetic moment scaling is applied relative to Jupiter, only the relative value of
the conductivity, i.e.~$\sigma/\sigma\J$ is required.
In this work, the conductivity is assumed to be the same for extrasolar gas giants as 
for Jupiter, i.e.~$\sigma/\sigma\J=1$. 

\paragraph{The size of the magnetosphere}

The size of the planetary magnetosphere $R\s$ is calculated with the parameters determined above for
the stellar wind and the planetary magnetic moment.
For a given planetary orbital distance $d$, of the different pressure contributions  
only the magnetospheric magnetic pressure is a function of 
the distance to the planet. Thus, the standoff distance $R_s$ is found from the pressure 
equilibrium
\citep{Griessmeier05}:
\begin{equation}
   R_s(d) = 
   \left[ \frac{\mu_0f_0^2\mathcal{M}^2}
   {8\pi^2 \left(m n(d) v\eff(d)^2+2 \, n(d)k_BT\right)} \right]^{1/6}.
   \label{eq:Rs}
\end{equation}
Here, $m$ is the proton's mass, and $f_0$ is the form factor of the magnetosphere. It describes the magnetic field 
created by the magnetopause currents. 
For a realistic magnetopause shape, a factor $f_0 = 1.16$ is used \citep{Voigt95}.
In term of units normalized to Jupiter's units, this is equivalent to
\begin{equation}
   R_s(d) \approx 40 R\J
   \left[ \frac{\widetilde{\mathcal{M}}^2}
   { \tilde{n}(d) \widetilde{v\eff}(d)^2+\frac{2 \, \tilde{n}(d)k_BT}{m {v_\text{eff,J}}^2}} \right]^{1/6},
\end{equation}
with $\tilde{n}(d)=n(d)/n\J$, $\widetilde{v\eff}(d)=v\eff(d)/v_\text{eff,J}$, $n\J=2.0 \cdot10^5$ m$^{-3}$,
and $v_\text{eff,J}=520$ km/s.

Note that in a few cases, especially for planets with very weak magnetic moments and/or subject
to dense and fast stellar winds of young stars, eq.~(\ref{eq:Rs}) yields standoff distances
$R_s<R\p$, where $R\p$ is the planetary radius.
Because the magnetosphere cannot be compressed to sizes smaller than the planetary radius, we 
set $R_s=R\p$ in those cases. 
 
\subsection{Additional parameters}

The other required parameters are obtained from the following sources:

\begin{itemize}
	\item The stellar ages $t_\star$ are taken from \citet[][who gave ages for 112 exoplanet host 
		stars]{Saffe05}.
		\citet{Saffe05} compare stellar age estimations based on five different methods (and for one of 
		them they use two different calibrations), 
		some of which give more 
		reliable results than others. Because not all of these
		methods are applicable to all stars, 
		we use the age
		estimations in the following order or preference: chromospheric age for ages 
		below 5.6 Gyr (using the D93 calibration), isochrone age, chromospheric age for 
		ages above 5.6 Gyr (using the D93 calibration), metallicity age.
		Note that error estimates for the two most reliable methods, namely isochrone 
		and chromospheric ages, are already relatively large (30\%-50\%).
 		In those cases where the age is not known, a default value of 5.2 Gyr is used 
		\citep[the median chromospheric age found by][]{Saffe05}.
		This relatively high average age is due to a selection effect (planet detection 
		by radial velocity method is easier to achieve for older, more slowly rotating 
		stars). As the uncertainty of the radio flux estimation becomes very large for 
		low stellar ages \citep{Griessmeier05}, we use a minimum stellar age of 0.5 Gyr.
	\item	For the solid angle of the beam, we assume the emission to be analogous to the 
		dominating contributions of Jupiter's radio emission and use $\Omega= 1.6$ sr 
		\citep{Zarka04}. 	
\end{itemize}

\section{Expected radio flux for know exoplanets} \label{sec:results}

\subsection{The list of known exoplanets} \label{sec:table}

Table 1 \footnote{Table 1 is only available in electronic form at the CDS.} shows what radio emission we expect from the presently known exoplanets (13.1.2007).
It contains the maximum emission frequency according to eq.~(\ref{eq:f}), 
the plasma frequency in the stellar wind at the planetary location according to eq.~(\ref{eq:fplasma})
and the expected radio flux according to the magnetic model (\ref{eq:Pin:kin}), the kinetic model (\ref{eq:Pin:mag}), and the kinetic CME model (\ref{eq:Pin:kin:CME}). 
The unipolar interaction model is discussed in the text below.
Table 1 also contains values for the expected planetary mass $M\p$, its radius $R\p$ and its planetary magnetic dipole moment $\mathcal{M}$. For each planet, we note whether tidal locking should be expected.

Note that  {\tt http://exoplanet.eu} contains a few more planets than table 1, because for some planets, essential data required for the radio flux estimation are not available (typically $s$, the distance to the observer).


The numbers given in table 1 are not accurate results, but should be
regarded as refined estimations intended to guide observations. Still, the errors and uncertainties involved in these
estimations can be considerable. As was shown in \citet{GriessmeierPSS06}, the uncertainty on the
radio flux at Earth, $\Phi$, is dominated by the uncertainty in the stellar age $t\sstar$
\citep[for which the error is estimated as $\approx 50$\% by][]{Saffe05}. For the maximum
emission frequency, $f\cc^{\text{max}}$, the error is determined by the uncertainty in the 
planetary magnetic moment $\mathcal{M}$, which is uncertain by a factor of 
two. For the planet $\tau$ Bootes b, these effects translate into an uncertainty of almost one order of magnitude for the flux (the error is smaller for planets around stars of solar age), 
and an uncertainty of a factor of 2-3 for the maximum emission frequency. This error estimate is derived and discussed in more detail by \citet{GriessmeierPSS06}.

The results given in table 1 cover the following range:
\begin{itemize}
	\item	The maximum emission frequency found to lie beween 0 to almost 200 MHz. However, all planets 
		with $ f_{\text{c}}^{\text{max}} > 70$ MHz have negligible flux. 
		Also note that any emission with $ f_{\text{c}}  \le 5$ to $10$ MHz
		will not be detectable on Earth because it cannot propagate through the Earth's ionosphere 
		(``ionospheric cutoff''). For this reason, the most appropriate frequency window for radio observations
		seems to be between 10 and 70 MHz.
	\item The radio flux according to the \textit{magnetic energy} model, $ \Phi_{\text{sw,mag}} $ lies between 0 and 
		5 Jy (for GJ 436 b). For 15 candidates, $ \Phi_{\text{sw,mag}} $ is larger than 100 mJy, and for 37 
		candidates it is above 10 mJy. 
	\item The flux prediction according to the \textit{kinetic energy}
		model, $ \Phi_{\text{sw,kin}} $ is much lower than the flux 
		according to the magnetic energy model. Only in one case it exceeds the value of 10 mJy.
	\item The increased stellar wind density and velocity during a \textit{CME} leads to a strong increase of the radio 
		flux when compared to the \textit{kinetic energy} model. Correspondingly, $ \Phi_{\text{CME,kin}} $ 
		exceeds 100 mJy in 3 cases and 10 mJy in 11 cases.
	\item	Table 1 does not contain flux estimations for the \textit{unipolar interaction} model. 						The reason is that  the condition given by eq.~(\ref{eq:f:generation}) is not satisfied in \textit{any} of the 
		studied cases.
		This is consistent with the result of \citet{Zarka06PREVI,Zarka06PSS}, who found that stars 	
		100 times as strongly magnetised as the sun are required for this type of emission. The approach taken in 
		section \ref{sec:stellarwind} for the estimation stellar magnetic fields does not yield such strong 
		magnetic fields for stars with ages $>0.5$ Gyr. 
		Stronger magnetic fields are possible (e.g.~for younger stars).
		Strongly magnetised stars (even those without known planets) could be defined as specific targets
		to test this model. 
	\item The plasma frequency in the stellar wind, $ f_{\text{p,sw}} $, is negligibly small in most cases. For a 
		few planets, however, it is of the same order of magnitude as the maximum emission frequency.
		In these cases, the condition given by eq.~(\ref{eq:escape}) 
		makes the escape of the radio emission from its source towards the observer impossible.
		Of the 197 planets of the current census, eq.~(\ref{eq:escape}) is violated in 8 cases. If one takes into 
		account the uncertainty of the stellar age \citep[30-50\%, see][]{Saffe05}, an uncertainty of similar size is 	
		introduced for the plasma frequency: In the example of $\tau$ Bootis, the age uncertainty 		
		translates into a variation of up to 50\% for the plasma frequency.  With such error bars, 								between 6 and 14 planets are affected by eq.~(\ref{eq:escape}). None of the best targets are 
		affected.
	\item	The expected planetary magnetic dipole moments lie between 0 and 5.5 times the magnetic moment 
		of Jupiter. However, the highest values are found only for very massive planets: Planets with masses 
		$M\le2M\J$ 
		have magnetic moments $\mathcal{M}\le2\mathcal{M}\J$. For planets with masses of $M\le M\J$, the 
		models predict magnetic moments $\mathcal{M}\le1.1\mathcal{M}\J$. 
\end{itemize}

The results of table 1 confirm that the different models for planetary radio emission lead to very different results. The largest fluxes are found for the \textit{magnetic energy} model, followed by the \textit{CME} model and the \textit{kinetic energy} model. This is consistent with previous expectations 
\citep{Zarka01,Zarka06PREVI,GriessmeierPREVI,Zarka06PSS,GriessmeierPSS06}.
The \textit{unipolar interaction} model does not lead to observable emission for the presently known exoplanets. Furthermore, the impact of tidal locking is clearly visible in the results. 
As it is currently not clear which of these models best describes the auroral radio emission, it is not sufficient to restrict oneself to one scaling law (e.g.~the one yielding the largest radio flux). For this reason, all possible  models have to be considered. Once exoplanetary radio emission is detected, observations will be used to constrain and improve the models.  

Table 1 also shows that planets subject to tidal locking have a smaller magnetic moment and thus a lower maximum emission frequency than freely rotating planets. The reduced bandwidth of the emission can lead to an increase of the radio flux, but frequently emission is limited to frequencies not observable on earth (i.e.~below the ionospheric cutoff).

\begin{figure}[ht]
\begin{center}
\centerline{\includegraphics[width=1.0\linewidth]{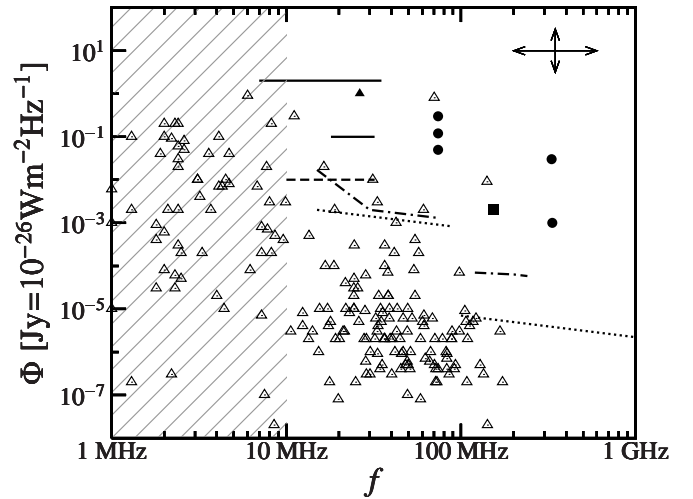}}
  \caption{
Maximum emission frequency and expected radio flux for known extrasolar planets according to the \textit{magnetic energy} model, compared to the limits of past and planned observation attempts. Open triangles: Predictions for planets. 
Solid lines and filled circles: Previous observation attempts at the UTR-2 (solid lines), at Clark Lake (filled triangle), at the VLA (filled circles), and at the GMRT (filled rectangle). For comparison, the expected sensitivity of new detectors is shown: upgraded UTR-2 (dashed line), 
LOFAR (dash-dotted lines, one for the low band and one for the high band antenna), LWA (left dotted line) and SKA (right dotted line). Frequencies below $~10$ MHz are not observable from the ground (ionospheric cutoff).  Typical uncertainties are indicated by the arrows in the upper right corner.
}
\label{fig:radiopredictionSWmag}
\end{center}
\end{figure}

\begin{figure}[ht]
\begin{center}
\centerline{\includegraphics[width=1.0\linewidth]{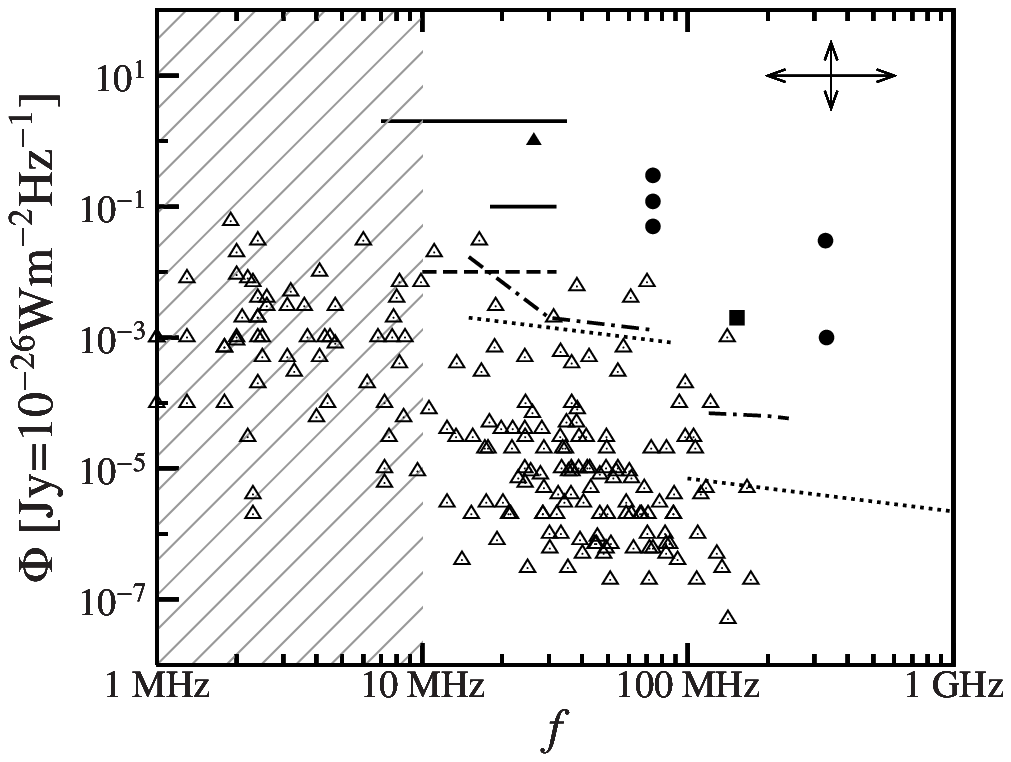}}
  \caption{Maximum emission frequency and expected radio flux for known extrasolar planets according to the \textit{CME} model, compared to the limits of past and planned observation attempts. Open triangles: Predictions for planets. All other lines and symbols are as defined in figure \ref{fig:radiopredictionSWmag}.}
\label{fig:radiopredictionCMEkin}
\end{center}
\end{figure}

\begin{figure}[ht]
\begin{center}
\centerline{\includegraphics[width=1.0\linewidth]{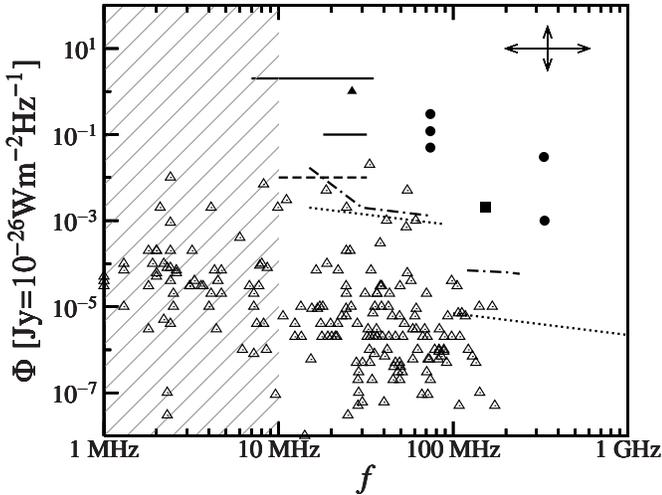}}
  \caption{Maximum emission frequency and expected radio flux for known extrasolar planets according to the \textit{kinetic energy} model, compared to the limits of past and planned observation attempts. Open triangles: Predictions for planets. 
All other lines and symbols are as defined in figure \ref{fig:radiopredictionSWmag}.}
\label{fig:radiopredictionSWkin}
\end{center}
\end{figure}

The results of table 1 are visualized in 
figure \ref{fig:radiopredictionSWmag} (for the \textit{magnetic energy} model),
figure \ref{fig:radiopredictionCMEkin} (for the \textit{CME} model) and
figure \ref{fig:radiopredictionSWkin} (for the \textit{kinetic energy} model).
The predicted planetary radio emission is denoted by open triangles (two for each ``potentially locked'' planet, 
otherwise one per planet). The typical uncertainties (approx.~one order of magnitude for the flux, 
and a factor of 2-3 for the maximum emission frequency) are indicated by the arrows in the upper right corner.
The sensitivity limit of previous observation attempts are shown as filled symbols and as solid lines \citep[a more detailed comparison of these observations can be found in][]{Zarka04a,Griessmeier51PEG05,GriessmeierPHD06}.  
The expected sensitivity of new and future detectors (for 1 hour integration and 4 MHz bandwidth, or any equivalent combination) is shown for comparison. Dashed line: upgraded UTR-2, dash-dotted lines: low band and high band of LOFAR, left dotted line: LWA, right dotted line: SKA. The instruments' sensitivities are defined by the radio sky background.
For a given instrument, a planet is observable if it is located
either above the instrument's symbol or above and to its right. Again, large differences in expected flux densities are apparent between the different models. On average, the \textit{magnetic energy} model yields the largest flux densities, and the  \textit{kinetic energy} model yields the lowest values.
Depending on the model, between one and three planets are likely to be observable using the upgraded system of UTR-2. 
Somewhat higher numbers are found for LOFAR.
Considering the uncertainties mentioned above, these numbers should not be taken literally, but should be seen as an indicator that while observation seem feasible, the number of suitable candidates is rather low. 
It can be seen that the maximum emission frequency of many planets lies below the ionospheric cutoff frequency, making earth-based observation of these planets impossible. A moon-based radio telescope however would give access to radio emission with frequencies of a few MHz \citep{Zarka06PSS}. As can be seen in figures  \ref{fig:radiopredictionSWmag},  
\ref{fig:radiopredictionCMEkin} and
\ref{fig:radiopredictionSWkin}, this frequency range includes a significant number of potential target planets with relatively high flux densities.

Figures \ref{fig:radiopredictionSWmag},  
\ref{fig:radiopredictionCMEkin} and
\ref{fig:radiopredictionSWkin} also show that 
the relatively high frequencies of the LOFAR high band and of the SKA telescope are probably not very well suited for the search for exoplanetary radio emission.
These instruments could, however, be used to search for radio emission generated by 
 \textit{unipolar interaction} between planets and strongly magnetised stars.  

\subsection{A few selected cases} \label{sec:cases}

According to our analysis, the best candidates are:
\begin{itemize}
	\item HD 41004 B b, which is the best case in the \textit{magnetic energy} model with emission above 1 MHz.
		Note that the mass of this object is higher than the upper limit for planets ($\approx 13 M\J$), so that it 
		probably is a brown dwarf and not a planet.
	\item Epsilon Eridani b, which is the best case in the \textit{kinetic energy} model.
	\item	Tau Boo b, which is the best case in the \textit{magnetic energy} model with emission above the 
		ionospheric cutoff (10 MHz).
	\item HD 189733 b, which is the best case in both the \textit{magnetic energy} model and in the 
		\textit{CME} model 
		which has emission above 1 MHz.  
	\item	Gliese 876 c, which is the best case in the \textit{CME} model with emission above the ionospheric cutoff 
		(10 MHz).
	\item	HD 73256 b, which has emission above 100 mJy in the \textit{magnetic energy} model and which is the 
		second best planet in the \textit{kinetic energy} model.
	\item	GJ 3021 b, which is the third best planet in the \textit{kinetic energy} model.
\end{itemize}

To this list, one should add the planets around Ups And (b, c and d) and of HD 179949 b, 
whose parent stars exhibit an increase of the chromospheric emission of about 
1-2\% \citep{Shkolnik03,Shkolnik04,Shkolnik05}.
The observations indicate one maximum per planetary orbit, 
a ``Hot Spot'' in the stellar chromosphere which is in phase with the 
planetary orbit.
The lead angles observed by 
\citet{Shkolnik03} and \citet{Shkolnik05} were recently explained with an Alfv\'{e}n-wing model
using realistic stellar wind parameters obtained from the stellar wind model by Weber and Davis 
\citep{PreussePHD05,Preusse06}. This indicates that a magnetised planet is not required to describe the
present data. 
The presence of a planetary magnetic field could, however, be proven by the existence of planetary radio emission.
Although our model does not predict high radio fluxes from these planets (see table 1), the high 
chromospheric flux shows that a strong interaction is taking place. As a possible solution of this problem, an intense stellar magnetic field was suggested \citep{Zarka06PREVI,Zarka06PSS}. In that case, table 1 underestimates the radio emission of Ups And b, c, d and HD 179949 b, making these planets interesting candidates for radio observations (eg.~through the \textit{magnetic energy} model or the \textit{unipolar interaction} model).
For this reason, it would be especially interesting to obtain measurements of the stellar magnetic field for these two
planet-hosting stars \citep[e.g.~by the method of][]{Catala07}.  
		
Considering the uncertainties mentioned above, it is important not to limit observations attempts to these best cases. The estimated radio characteristics should only be used as a guide (e.g.~for the target selection, or for statistical analysis), but individual results should not be regarded as precise values.

\subsection{Statistical discussion} \label{sec:statistics}

It may seem surprising that so few good candidates are found among the 197 examined exoplanets. However, when one checks the list of criteria for ``good'' candidates \citep[e.g.][]{Griessmeier51PEG05}, it is easily seen that only a few good targets can be expected: (a) the planet should be close to the Earth (otherwise the received flux is too weak). About 70\% of the known exoplanets are located within 50 pc, so that this is not a strong restriction. (b) A strongly magnetised system is required (especially for frequencies above the ionospheric cutoff). For this reason, the planet should be massive (as seen above, we find magnetic moments 
$\mathcal{M}\ge2\mathcal{M}\J$ only for planets with masses $M\ge2M\J$). About 60\% of the known exoplanets are at least as massive as Jupiter (but only 40\% have $M\p \ge 2.0 M\J$). (c) The planet should be located close to its host star to allow for strong interaction (dense stellar wind, strong stellar magnetic field). Only 25\% of the known exoplanets are located within 0.1 AU of their host star. 

By multiplying these probabilities, one finds that close ($s \le 50$ pc), heavy
($M\p \ge 2.0 M\J$), close-in  ($d \le 0.1$ AU) planets would represent 8\% ($\approx$ 15 of 197 planets) of the current total if the probabilities for the three conditions were independent. However, this is not the case. In the current census of exoplanets, a correlation between planetary mass and orbital distance is clearly evident, with a lack of close-in massive planets \citep[see e.g.][]{Udry03}. This is not a selection effect, as massive close-in planets should be easier to detect than low-mass planets. This correlation was explained by the stronger tidal interaction effects for massive planets, leading to a faster decrease of the planetary orbital radius until the planet reaches the stellar Roche limit and is effectively destroyed \citep{Paetzold02,Jiang03}. Because of this mass-orbit correlation, the fraction of good candidates is somewhat lower (approx.~2\%, namely 3 of 197 planets:  HD 41004 B b, Tau Boo b, and HD 162020 b).

\subsection{Comparison to previous results} \label{sec:comparison}

A first comparative study of expected exoplanetary radio emission from a large number of planets was performed by \citet{Lazio04}, who compared expected radio fluxes of 118 planets (i.e.~those known as of 2003, July 1).
Their results differ considerably from those given in table 1:
\begin{itemize}
	\item	As far as the maximum emission frequency $f_\text{c}^{\text{max}}$ is concerned, 
		our results are considerably lower than the frequencies given by 
		\citet{Lazio04}.
		For Tau Bootes, their maximum emission frequency is six times larger than our result.
		For planets heavier than Tau Bootes, the discrepancy is even larger, reaching more than 
		one order of 
		magnitude for the very heavy cases (e.g.~HD 168433c, for which they predict radio emission 
		with frequencies up to 2670 MHz). 
		These differences have several reasons: Firstly, \citet{Farrell99} and \citet{Lazio04} assume 
		that $R\p=R\J$. Also, these works rely on the magnetic 
		moment scaling law of \citet{Blackett47}, which has a large exponent in $r\cc$. 
		This scaling law should not be used, as it was experimentally disproven \citep{Blackett52}.
		Thirdly, these works make uses $r\cc\propto M\p^{1/3}$. 
		Especially for planets with large masses like $\tau$ Bootes, this yields 
		unrealistically large core radii (even $r\cc>R\p$ in some cases), magnetic moments, 
		and emission frequencies.
		Note that a good estimation of the emission frequency is particularly important,
		because a difference of a factor of a few can make the difference between 
		radiosignals above and below the Earth's ionospheric cutoff frequency.
	\item	The anticipated radio flux obtained with the \textit{kinetic energy} model $\Phi_{\text{sw,kin}}$ 
		is much lower than the estimates of \citet{Lazio04}.
		Typically, the difference is approximately two orders of magnitude, but this varies strongly 
		from case 
		to case. For example, for Tau Boo b, our result is smaller by a factor of 30 (where the 
		difference is partially 
		compensated by the low stellar age which increases our estimation), for Ups And b, 
		the results differ by a factor of 220, and for Gliese 876 c, 
		the difference is as large as a factor 6300 (which is partially due to the fact we take into 
		account the
		small stellar radius and the high stellar age).
	\item	
		As was mentioned above, the analysis of the jovian radio emission allows to define three 
		terms for the typical radio spectrum:  (a) the power during \textit{average conditions}, 
		(b) the average power during 
		periods of \textit{high activity}, and (c) the \textit{peak power} \citep{Zarka04}.
		When comparing the results of our table 1 to those of 
		\citet{Lazio04}, one has to note that their table I gives the 
		\textit{peak power}, while the results in our table 1 were 
		obtained using the average power during periods of \textit{high activity}.
		Similarly to \citet{Farrell99}, \citet{Lazio04} assume that the peak power 
		caused by variations of the stellar wind velocity is two orders of magnitude 
		higher than the average power.
		However, 	the values \citet{Farrell99} use for average conditions correspond to
		periods of high activity, which are less than one order of magnitude below 
		the peak power \citep[][]{Zarka04}. 
		During periods of peak emission, 
		the value given in our table 1 would be increased by the same
		amount \citep[approximately a factor of 5, see][]{Zarka04}.
		For this reason, 
		the peak radio flux is considerably overestimated in these studies.
	\item	Estimated radio fluxes according to the \textit{magnetic energy} model and the \textit{CME} 
		model have not yet been published for large numbers of planets. This is the first time the results from 
		these models are compared for a large number of planets.
\end{itemize}

\section{Conclusions} \label{sec:conclusion}

Predictions concerning the radio emission from all presently known extrasolar planets were presented.
The main parameters related to such an emission were analyzed, namely the planetary magnetic dipole moments, the maximum frequency of the radio emission, the radio flux densities, and the possible escape of the radiation towards a remote observer. 

We compared the results obtained with various theoretical models.
Our results confirm that the four different models for planetary radio emission lead to very different results. As expected, the largest fluxes are found for the \textit{magnetic} energy model, followed by the \textit{CME} model and the \textit{kinetic} energy model. The results obtained by the latter model are found to be less optimistic than by previous studies. The \textit{unipolar interaction} model does not lead to observable emission for any of the currently known planets. As it is currently not clear which of these models best describes the auroral radio emission, it is not sufficient to restrict oneself to one scaling law (e.g.~the one yielding the largest radio flux). Once exoplanetary radio emission is detected, observations will be used to constrain and improve the model.  

These results will be particularly useful for the target selection of current and future radio observation campaigns
(e.g.~with the VLA, GMRT, UTR-2 and with LOFAR).  We have shown that observation seem feasible, but that the number of suitable candidates is relatively low. The best candidates appear to be HD 41004 B b, Epsilon Eridani b, Tau Boo b, HD 189733 b, Gliese 876 c, HD 73256 b, and GJ 3021 b. The observation of some of these candidates is in progress.

\begin{acknowledgements}
We thank J. Schneider for providing data via ``The extrasolar planet encyclopedia'' (http://exoplanet.eu/), I. Baraffe and C. Vocks for helpful discussions concerning planetary radii. 
We would also like to thank the anonymous referee for his helpful comments. 
This study was jointly performed within the ANR
project ``La d\'{e}tection directe des exoplan\`{e}tes en ondes radio'' and within the
LOFAR transients key project (TKP). 
 J.-M. G. was supported by the french national research agency (ANR) within the project with 
the contract number NT05-1\_42530 and partially by Europlanet (N3 activity). 
P.Z. acknowledges support from the International Space Science Institute (ISSI) within the ISSI team ``Search for Radio Emissions from Extra-Solar Planets''. 
\end{acknowledgements}

\begin{appendix}
\section{An empirical mass-radius relation}\label{sec:appendix}

For the selection of targets for the search for radio emission from extrasolar planets, an
estimation of the expected radio flux $\Phi$ and of the maximum emission frequency 
$f\cc^{\text{max}}$ is required. For the calculation of these values, both the planetary mass 
and the planetary radius are required
\citep[see, e.g.~][]{Farrell99,GriessmeierPSS06,Zarka06PSS}.
However, only for a few planets (i.e. the 16 presently known transiting planets) both mass and radius are known.

In the absence of observational data, it is in principle possible to obtain planetary radii from numerical simulation, e.g.~similar to those of \citet{Bodenheimer03} or \citet{Baraffe03,Baraffe05}, requiring one numerical run per planet. We chose instead to derive a simplified analytical fit to such numerical results.

\subsection{The accuracy of the fit}
\label{sec-precision}

The description presented here is necessarily only preliminary, as (a) numerical models 
are steadily further developed and improved, and as (b) more transit observations 
(e.g.~by the COROT satellite, which was launched recently) will 
provide a much better database in the future.
This will considerably improve our understanding of the dependence of the planetary radius on 
various parameters as, e.g.~planetary mass, orbital distance, or stellar  metallicity 
\citep[as suggested by][]{Guillot06}.

\subsubsection{What accuracy can we accept?}

Within the frame of the models presented in section \ref{sec:theory}, an increase in $R\p$ by 40\% increases the expected radio flux by a factor of 2, and the estimated maximum emission frequency decreases by 40\%. 
More generally, for a fixed planetary mass, $\Phi$ is roughly proportional to $R\p^{7/3}$ and 
$f\cc^{\text{max}}$ is approximately proportional to $R\p^{-1}$.
Thus, it appears that the assumption of a single standard radius for all planets leads to a relatively large error. 
Comparing this to the other uncertainties involved in the estimation of radio characteristics 
\citep[these are discussed in][]{GriessmeierPSS06}, it seems sufficient to estimate $R\p$ with 20\% accuracy.

\subsubsection{What accuracy can we expect?}

Several effects limit the precision in planetary radius we can hope to achieve:
\begin{itemize}
	\item	The definition of the radius: 
		The ``transit radius'' measured for transiting planets 
		is not exactly identical to the standard 1 bar radius.
		The differences are of the order of about 5-10\%, but depend on the mass of the planet
		\citep{Burrows03,Burrows04}.
		Because we compare modelled radii and observed radii without correcting for this effect, 
		this limits the maximum precision we can potentially obtain. 
	\item The abundance of heavy elements:
		The transiting planet around HD 149026 is substantially enriched in heavy 
		elements \citep{Sato05}.
		Models \citep{Bodenheimer03} yield a smaller radius for planets with 
		a heavy core than for coreless planets of the same mass (more than 10\% difference for small 
		planets). 
		For a planet with unknown radius, a strong enrichment in heavy elements cannot 
		be ruled out, as this case cannot  
		be distinguished observationally from a pure hydrogen giant 
		(i.e.~one without heavy elements). 
\end{itemize}

For these reasons, we conclude that an analytical description which agrees with the (numerical) data within $\sim$20\% seems
sufficient. For such a description, the error introduced by the fit will not be the dominant one. To get a better result, it 
is not sufficient to improve the approximation for the radius estimation, but the more fundamental problems mentioned above have to be addressed.

\subsection{An analytical mass-radius relation}
\label{sec-model}

\subsubsection{The influence of the planetary mass}
\label{sec-model-mass}

A simple mass-radius relation, valid within a vast mass range, has been proposed by
\citet{LyndenBell01preprint} and \citet{LyndenBell01}:
\begin{equation}
	R\p=
	\frac{1}{\left( \frac{4}{3} \pi \rho_0 \right)^{1/3}} 
	\frac{M\p^{1/3}}{1+\left( \frac{M\p}{M_{\text{max}}} \right)^{2/3}}
	\label{eq:mass-radius}
\end{equation}
The density $\rho_0$ is depends on the planetary atomic composition.
Eq.~(\ref{eq:mass-radius}) has a maximum in $R\p$ when $M\p=M_{\text{max}}$ 
(corresponding to the planet of maximum radius).
Fitting Jupiter, Saturn and the planet of maximum radius \citep[$R_{\text{max}}=1.16 R\J$, see][]{Hubbard84}, we obtain $M_{\text{max}}=3.16 M\J$ and 
$\rho_0=394$ kg m$^{-3}$ for a Jupiter-like mixture of hydrogen and helium (75\% and 25\% by mass, respectively). For a  pure hydrogen planet, $\rho_0=345$ kg m$^{-3}$.
With there parameters, eq.~(\ref{eq:mass-radius}) fits well the results for cold planets of 
\citet{Bodenheimer03}. 
In this work, this estimation is used for the radii of non-irradiated (cold) planets.

\subsubsection{The influence of the planetary age}
\label{sec-model-age}

It is known that the radius of a planet with a given mass depends on its age.
According to models for the radii of isolated planets \citep{Baraffe03},
the assumption of a time-independent planetary radius leads to an error $\le$11\% for planets with ages above 0.5 Gyr and with masses above $0.5 M\J$. 
In view of the uncertainties discussed above, this error seems acceptable for a first approximation, and we use
\begin{equation}
	R_{\text{p}}(M\p,t)\approx R_{\text{p}}(M\p) \label{eq-Riso}.
\end{equation}

\subsubsection{The influence of the planetary orbital distance}
\label{sec-model-distance}

It is commonly expected that planets subjected to strong stellar radiation have a larger planetary radius than isolated, but otherwise identical planets. 
This situation is typical for ``Hot Jupiters'', where strong stellar irradiation is supposed
to delay the planetary contraction
\citep{Burrows00,Burrows03,Burrows04}.  
Clearly, this effect depends on the planetary distance to its star, $d$, and on the stellar luminosity
$L\sstar$.
In the following, we denote the radius increase by $r$, which we define as 
\begin{equation}
	r=\frac{R\p(M\p,d)}{R\p(M\p,d=\infty)}.
\end{equation}
Herein, $R\p(M\p,t,d)$ denotes the radius of a planet under the irradiation by its host star, and $R\p(M\p,t,d=\infty)$ is the radius of a non-irraditated, but otherwise identical planet. 

Different exoplanets have vastly different host stars. A difference of a factor of two in stellar mass can result in a difference of more than order of magnitude in stellar luminosity. For this reason, it is not sufficient to take the orbital distance as the only parameter determining the radius increase by irradiation.
 
Here, we select the equilibrium temperature of the planetary surface as the basic parameter. 
It is defined as 
\citep{Bodenheimer03}
\begin{equation}
	T_{\text{eq}}=\left[
			\frac{(1-A)L\sstar}{16 \pi \sigma_{\text{SB}} d^2 (1+e^2/2)^2}
			\right]^{1/4},
\end{equation}
where the planetary albedo is set as $A=0.4$ in the following. The stellar luminosity $L\sstar$ is calculated from the stellar mass according to the analytical fit given by \citet{Tout96}
for the zero-age main sequence. In this, we assumed solar metallicity for all stars. 
Finally, $\sigma_{\text{SB}}$ denotes the Stefan-Boltzmann constant.

Using the data of \citet{Bodenheimer03}, we use the following fit for the radius increase $r$:
\begin{equation}
	r=1+0.05\left(\frac{T_{\text{eq}}}{T_0}\right)^{\gamma}. \label{eq:rratio}
\end{equation}
This form was selected because it has the correct qualitative behaviour: It yields a monotonous decrease of $r$ with decreasing equilibrium temperature $T_{\text{eq}}$ 
(increasing orbital distance $d$). 
In the limit $T_{\text{eq}}\to0$ ($d\to\infty$) we find $r\to 1$. 
As an alternative to a continuous fit like eq.~(\ref{eq:rratio}),
\citet{Lazio04} use a step-function, i.e.~assume an increased radius of $r=1.25$ 
for planets closer than a certain distance $d_1=0.1$ AU only. 
However, this treatment  
does not reduce the number of fit-constants.
Also, for planets with orbital distances close to $d_1$, 
the results obtained with a step-function strongly depend on the somehow arbitrary choice of $d_1$. A continuous transition from ``irradiated'' to ``isolated'' planets is less prone to this effect.

\begin{figure}[ht]
\begin{center}
\centerline{\includegraphics[width=1.0\linewidth]{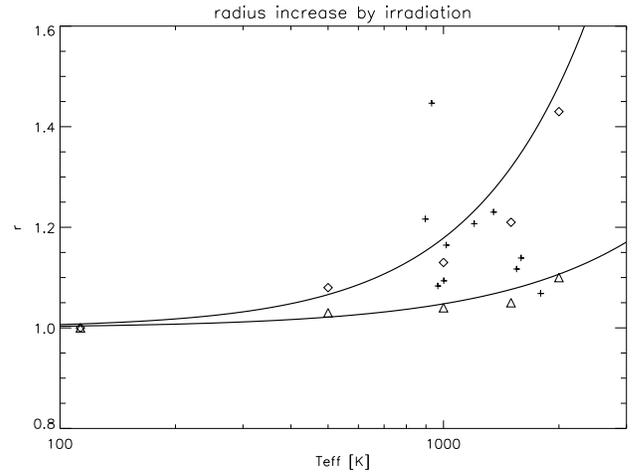}}
  \caption{Radius increase $r$ as a function of the equilibrium temperature $T_{\text{eq}}$ 
  according to eq.~(\ref{eq:rratio}), (\ref{eq:T0}) and (\ref{eq:gamma}). Upper line: radius increase $r$ due to irradiation for planets of mass $M\p=0.11M\J$. Lower line: Upper line: radius increase $r$ due to irradiation for planets of mass $M\p=3.0M\J$. Open symbols: results of the numerical calculation of 
\citet{Bodenheimer03} for planetary masses $M\p=0.11M\J$ (diamonds) and $M\p=3.0M\J$ (triangles).
Crosses: radius increase for observed transiting planets (see text).
}
\label{fig:rofTeff}
\end{center}
\end{figure}

The numerical results of \citet{Bodenheimer03} show that the ratio $r$ also depends on the mass of the planet: for small planets, $r=1.4$ for an equilibrium temperature of 2000 K, whereas for large planets, $r\le 1.10$. For this reason, we allow to coefficients $T_0$ and $\gamma$ to vary with $M\p$:
\begin{equation}
	T_0=c_{\text{t,1}}\cdot \left( \frac{M\p}{M\J}\right)^{c_{\text{t,2}}}\label{eq:T0}
\end{equation}
and
\begin{equation}
	\gamma=1.15+0.05 \cdot \left(  \frac{c_{\gamma,1}}  {M\p} \right)^{c_{\gamma,2}}.\label{eq:gamma}
\end{equation}
With the set of coefficient $c_{\text{t,1}}=764$ K , $c_{\text{t,2}}=0.28$, $c_{\gamma,1}=0.59 M\J$ and $c_{\gamma,2}=1.03$, we obtain an analytical fit to the numerical results of \citet{Bodenheimer03}. The maximum deviation from the numerical results is below 10\% (cf.~Fig. \ref{fig:rofTeff}).

For comparison, Fig.~\ref{fig:rofTeff} also shows the value of $r$ for the transiting exoplanets
OGLE-TR-10b, OGLE-TR-56b, OGLE-TR-111b, OGLE-TR-113b, OGLE-TR-132b, XO-1b, HD 189733b, HD 209458b, TrES-1b and TrES-2b
as small crosses. Here, $r$ is calculated as the ratio of the observed value of $R\p(M\p,d)$ and the value for  $R\p(M\p,d=\infty)$ calculated according to eq.~(\ref{eq:mass-radius}). As the mass of all transiting planets lies between $0.11M\J \le M\p \le 3.0 M\J$, one should expect to find all crosses between the two limiting curves. For most planets, this is indeed the case: only for HD 209458b, $r$ is considerably outside the area delimited by the two curves. Different explanations have been put forward for the anomalously large radius of this planet, but so far no conclusive answer to this question has been found \citep[see e.g.][and references therein]{Guillot06}. As numerical models cannot reproduce the observed radius of this planet, one cannot expect our approach (which is based on a fit to numerical results) to reproduce it either. For the other planets, Fig.~\ref{fig:rofTeff} shows that our approach is a valid approximation.

\end{appendix}

\begin{appendix}
\section{Tidal locking?} \label{sec:tlocking}

For the estimation of the planetary magnetic dipole moment in section \ref{sec:protation}, 
we require the planetary rotation rate. This rotation rate rotation greatly depends on whether the planet can be considered as tidally locked, as freely rotating, or as potentially locked. In this appendix, we discuss how we evaluate 
the tidal locking timescale $\tis$, which decides to which of these categories a planet belongs.

The value of the planetary rotation $\omega$ depends on its distance from the central star.
For close-in planets, the planetary rotation rate is reduced by tidal dissipation.
In this case, tidal interaction gradually slows down the planetary rotation from its initial 
value $\omega_\text{i}$ until it reaches the final value $\omega_\text{f}$ after tidal locking 
is completed.

It should be noted that for planets in eccentric orbits, tidal interaction does not
lead to the synchronisation of the planetary rotation with the orbital period. Instead, the 
rotation period also depends on the orbital eccentricity. At the same time, the timescale to 
reach this equilibrium rotation rate is reduced \citep{Laskar04}. 
Similarly, for planets in an oblique orbit, the equilibrium rotation period is modified 
\citep{Levrard07}.
However, the locking of a hot Jupiter in a non-synchronous spin-orbit resonance appears to be unlikely
for distances $\leq$  0.1 AU \citep{Levrard07}.
For this reason, we will calculate the timescale for tidal locking under the assumptions of 
circular orbits and zero obliquity.
In the following, the tidal locking timescale for reaching $\omega_\text{f}$ is calculated under 
the following simplifying assumptions: prograde orbit, spin parallel to orbit (i.e.~zero 
obliquity), and zero eccentricity \citep[][Chapter 4]{Murray99}.

The rate of change of the planetary rotation velocity $\omega$ for a planet with a mass of 
$M\p$ and radius of $R\p$ around a star of mass $M\sstar$ 
is given by \citep{Goldreich66,Murray99}:
\begin{equation}
        \frac{d\omega}{dt}= \frac{9}{4} \frac{1}{\alpha \, Q\p'} \left( \frac{GM\p}{R\p^3} \right)
        \left( \frac{M\sstar}{M\p} \right)^2
        \left( \frac{R\p}{d} \right)^6
	,\label{eq:omegadot}
\end{equation}
where the constant $\alpha$ depends on the internal mass distribution within the
planet. 
It is 
defined by $\alpha=I/(M\p R\p^2)$,  
where $I$ is the planetary moment of inertia. 
For a sphere of homogeneous density, $\alpha$ is equal to 2/5. For planets, generally  
$\alpha\le 2/5$.
$Q\p'$ is the modified $Q$-value of the planet. 
It can be expressed as \citep{Murray99}
\begin{equation}
	Q\p'=\frac{3Q\p}{2k_{2,\text{p}}}, \label{eq:k_2}
\end{equation}	
where $k_{2,\text{p}}$ is the Love number of the planet.
$Q\p$ is the planetary tidal dissipation factor (the larger it is, the smaller is the tidal
dissipation), defined by \citet{MacDonald64} and \citet{Goldreich66}.

The time scale for tidal locking is obtained by a comparison of the planetary angular velocity
and its rate of change: 
\begin{equation}
	\tis=\frac{\omega_\text{i}-\omega_\text{f}}{\dot{\omega}}\,. \label{eq:tau}
\end{equation}
A planet with angular velocity $\omega_\text{i}$ at $t=0$ (i.e.~after formation) will gradually lose
angular momentum, until the angular velocity reaches $\omega_\text{f}$ at $t=\tis$.
Insertion of eq.~(\ref{eq:omegadot}) into eq.~(\ref{eq:tau}) yields the following expression for
$\tis$:
\begin{equation}
        \tis \approx \frac{4}{9} \alpha \, Q\p' \left( \frac{R\p^3}{GM\p} \right)
        \left( \omega_\text{i}-\omega_\text{f} \right)
        \left( \frac{M\p}{M\sstar} \right)^2
        \left( \frac{d}{R\p} \right)^6
	\label{eq:locking}
\end{equation}

The importance of this effect strongly depends on the distance ($\tis\propto d^{6}$). Thus, a 
planet in a close-in orbit ($d\lesssim 0.1$ AU) around its central star is subject to strong 
tidal interaction, leading to gravitational locking on a very short timescale. 

In the following, we briefly describe the parameters ($\alpha$, $Q\p'$, $\omega_\text{i}$ and 
$\omega_\text{f}$) required to calculate the timescale for tidal locking of Hot Jupiters.

\paragraph{Structure parameter $\alpha$}

For large gaseous planets, the equation of state can be approximated by a polytrope of index
$\kappa=1$. In that case, the structure parameter 
$\alpha$ (defined by $\alpha=I/M\p R\p^2$) is given by $\alpha=0.26$ 
\citep{Gu03}.

\paragraph{Tidal dissipation factor $Q_\mathrm{p}'$} \label{sec:Q':gas}

For planets with masses of the order of one Jupiter mass, one finds that
$k_{2,\text{p}}$ has a value of $k_{2,\text{p}}\approx 0.5$ for 
Jupiter \citep{Murray99,Laskar04} and $k_{2,\text{p}}\approx 0.3$ for Saturn \citep{Peale99,Laskar04}. 
The value of $k_{2,\text{p}}=0.5$ will be used in this work. 
With eq.~(\ref{eq:k_2}), this results in $Q\p'\approx 3 Q\p$.

For Jupiter, 
one finds
the following 
range of allowed values:
$6.6\cdot10^4\lesssim Q\p\lesssim2.0\cdot10^6$ \citep{Peale99}. 
Several estimations of the turbulent dissipation within Jupiter yield $Q\p$-values larger than 
this upper limit, while other theories predict values consistent with this upper limit
\citep[][and references therein]{Marcy97,Peale99}.
This demonstrates that the origin of the value of $Q\p$ is not well understood even for Jupiter
\citep{Marcy97}. 

Extrasolar giant planets are subject to strongly different conditions, and it is difficult to 
constrain $Q\p$.
Typically, Hot Jupiters are assumed to behave similarly to Jupiter, and values 
in the range
$1.0\cdot10^5 \le Q\p'  \le 1.0\cdot10^6$ are used. 
In the following, we will distinguish three different regimes: close planets (which are tidally locked), 
distant planets (which are freely rotating), and planets at intermediate distances (which are 
potentially tidally locked). The borders between the ``tidally locked'' and the ``potentially locked'' regime 
is calculated by setting $\tis=100$ Myr and $Q\p'=10^6$.
The border between the ``potentially locked'' and the ``freely rotating'' regime is 
calculated by setting $\tis=10$ Gyr and $Q\p'=10^5$.
Thus, the area of ``potentially locked'' planets is increased.

\paragraph{Initial rotation rate $\omega_\mathrm{i}$}

The initial rotation rate $\omega_\text{i}$ 
is not well constrained by planetary formation theories. 
The relation between the planetary angular momentum density and planetary mass observed in the 
solar system \citep{MacDonald64} suggests a primordial rotation period of the order of 
10 hours \citep[][Chapter 4]{Hubbard84}. 
We assume the initial rotation rate to be equal to the current rotation rate of Jupiter 
(i.e.~$\omega=\omega_\text{i}=\omega\J$) with $\omega\J=1.77\cdot10^{-4}$ s$^{-1}$.

\paragraph{Final rotation rate $\omega_\mathrm{f}$}

As far as eq.~(\ref{eq:locking}) is concerned, $\omega_\mathrm{f}$ can be neglected
\citep{GriessmeierPHD06}. 

\end{appendix}

\bibliographystyle{/home/griessmeier/tex/aa}

\begin{thebibliography}{80}
\expandafter\ifx\csname natexlab\endcsname\relax\def\natexlab#1{#1}\fi

\bibitem[{Baraffe {et~al.}(2003)Baraffe, Chabrier, Barman, Allard, \&
  Hauschildt}]{Baraffe03}
Baraffe, I., Chabrier, G., Barman, T.~S., Allard, F., \& Hauschildt, P.~H.
  2003, Astron. Astrophys., 402, 701

\bibitem[{Baraffe {et~al.}(2005)Baraffe, Chabrier, Barman, Selsis, Allard, \&
  Hauschildt}]{Baraffe05}
Baraffe, I., Chabrier, G., Barman, T.~S., {et~al.} 2005, Astron. Astrophys.,
  436, L47

\bibitem[{Blackett(1947)}]{Blackett47}
Blackett, P. M.~S. 1947, Nature, 159, 658

\bibitem[{Blackett(1952)}]{Blackett52}
Blackett, P. M.~S. 1952, Phil. Trans. R. Soc. A, 245, 309

\bibitem[{Bodenheimer {et~al.}(2003)Bodenheimer, Laughlin, \&
  Lin}]{Bodenheimer03}
Bodenheimer, P., Laughlin, G., \& Lin, D. N.~C. 2003, Astrophys. J., 592, 555

\bibitem[{Burrows {et~al.}(2000)Burrows, Guillot, Hubbard, Marley, Saumon,
  Lunine, \& Sudarsky}]{Burrows00}
Burrows, A., Guillot, T., Hubbard, W.~B., {et~al.} 2000, Astrophys. J., 534,
  L97

\bibitem[{Burrows {et~al.}(2004)Burrows, Hubeny, Hubbard, Sudarsky, \&
  Fortney}]{Burrows04}
Burrows, A., Hubeny, I., Hubbard, W.~B., Sudarsky, D., \& Fortney, I.~J. 2004,
  Astrophys. J., 610, L53

\bibitem[{Burrows {et~al.}(2003)Burrows, Sudarsky, \& Hubbard}]{Burrows03}
Burrows, A., Sudarsky, D., \& Hubbard, W.~B. 2003, Astrophys. J., 594, 545

\bibitem[{Cain {et~al.}(1995)Cain, Beaumont, Holter, Wang, \&
  Nevanlinna}]{Cain95}
Cain, J.~C., Beaumont, P., Holter, W., Wang, Z., \& Nevanlinna, H. 1995, J.
  Geophys. Res., 100, 9439

\bibitem[{Catala {et~al.}(2007)Catala, Donati, Shkolnik, Bohlender, \&
  Alecian}]{Catala07}
Catala, C., Donati, J.-F., Shkolnik, E., Bohlender, D., \& Alecian, E. 2007,
  Mon. Not. R. Astron. Soc., 374, L42

\bibitem[{Chandrasekhar(1957)}]{Chandrasekhar57}
Chandrasekhar, S. 1957, An Introduction to the Study of Stellar Structure,
  Dover Books on Astronomy and Astrophysics (New York: Dover Publications,
  Inc.)

\bibitem[{{Collier Cameron} \& Jianke(1994)}]{CollierCameron94}
{Collier Cameron}, A. \& Jianke, L. 1994, Mon. Not. R. Astron. Soc., 269, 1099

\bibitem[{Correia {et~al.}(2003)Correia, Laskar, \& de~Surgy}]{Correia03}
Correia, A. C.~M., Laskar, J., \& de~Surgy, O.~N. 2003, Icarus, 163, 1

\bibitem[{Desch \& Kaiser(1984)}]{Desch84}
Desch, M.~D. \& Kaiser, M.~L. 1984, Nature, 310, 755

\bibitem[{Dobbs-Dixon {et~al.}(2004)Dobbs-Dixon, Lin, \&
  Mardling}]{DobbsDixon04}
Dobbs-Dixon, I., Lin, D. N.~C., \& Mardling, R.~A. 2004, Astrophys. J., 610,
  464

\bibitem[{Farrell {et~al.}(1999)Farrell, Desch, \& Zarka}]{Farrell99}
Farrell, W.~M., Desch, M.~D., \& Zarka, P. 1999, J. Geophys. Res., 104, 14025

\bibitem[{Farrell {et~al.}(2004)Farrell, Lazio, Zarka, Bastian, Desch, \&
  Ryabov}]{Farrell04}
Farrell, W.~M., Lazio, T. J.~W., Zarka, P., {et~al.} 2004, Planet. Space Sci.,
  52, 1469

\bibitem[{Goldreich \& Soter(1966)}]{Goldreich66}
Goldreich, P. \& Soter, S. 1966, Icarus, 5, 375

\bibitem[{Grie{\ss}meier(2006)}]{GriessmeierPHD06}
Grie{\ss}meier, J.-M. 2006, PhD thesis, Technische Universit\"{a}t
  Braunschweig, {I}SBN 3-936586-49-7, {C}opernicus-GmbH Katlenburg-Lindau,
  {U}RL: http://www.digibib.tu-bs.de/?docid=00013336

\bibitem[{Grie{\ss}meier {et~al.}(2006{\natexlab{a}})Grie{\ss}meier,
  Motschmann, Glassmeier, Mann, \& Rucker}]{Griessmeier51PEG05}
Grie{\ss}meier, J.-M., Motschmann, U., Glassmeier, K.-H., Mann, G., \& Rucker,
  H.~O. 2006{\natexlab{a}}, in Tenth Anniversary of 51 Peg-b : Status of and
  Prospects for hot Jupiter studies, ed. L.~Arnold, F.~Bouchy, \& C.~Moutou
  (Platypus Press), 259--266, {U}RL:
  http://www.obs-hp.fr/www/pubs/Coll51Peg/proceedings.html

\bibitem[{Grie{\ss}meier {et~al.}(2006{\natexlab{b}})Grie{\ss}meier,
  Motschmann, Khodachenko, \& Rucker}]{GriessmeierPREVI}
Grie{\ss}meier, J.-M., Motschmann, U., Khodachenko, M., \& Rucker, H.~O.
  2006{\natexlab{b}}, in Planetary Radio Emissions VI, ed. H.~O. Rucker, W.~S.
  Kurth, \& G.~Mann (Austrian Academy of Sciences Press, Vienna), 571--579

\bibitem[{Grie{\ss}meier {et~al.}(2005)Grie{\ss}meier, Motschmann, Mann, \&
  Rucker}]{Griessmeier05}
Grie{\ss}meier, J.-M., Motschmann, U., Mann, G., \& Rucker, H.~O. 2005, Astron.
  Astrophys., 437, 717

\bibitem[{Grie{\ss}meier {et~al.}(2007)Grie{\ss}meier, Preusse, Khodachenko,
  Motschmann, Mann, \& Rucker}]{GriessmeierPSS06}
Grie{\ss}meier, J.-M., Preusse, S., Khodachenko, M., {et~al.} 2007, Planet.
  Space Sci., 55, 618

\bibitem[{Grie{\ss}meier {et~al.}(2004)Grie{\ss}meier, Stadelmann, Penz,
  Lammer, Selsis, Ribas, Guinan, Motschmann, Biernat, \& Weiss}]{Griessmeier04}
Grie{\ss}meier, J.-M., Stadelmann, A., Penz, T., {et~al.} 2004, Astron.
  Astrophys., 425, 753

\bibitem[{Gu {et~al.}(2003)Gu, Lin, \& Bodenheimer}]{Gu03}
Gu, P.-G., Lin, D. N.~C., \& Bodenheimer, P.~H. 2003, Astrophys. J., 588, 509

\bibitem[{Guillot {et~al.}(2006)Guillot, Santos, Pont, Iro, Melo, \&
  Ribas}]{Guillot06}
Guillot, T., Santos, N.~C., Pont, F., {et~al.} 2006, Astron. Astrophys., 453,
  L21

\bibitem[{Halbwachs {et~al.}(2005)Halbwachs, Mayor, \& Udry}]{Halbwachs05}
Halbwachs, J.~L., Mayor, M., \& Udry, S. 2005, Astron. Astrophys., 431, 1129

\bibitem[{Hilgers(1992)}]{Hilgers92}
Hilgers, A. 1992, Geophys. Res. Lett., 19, 237

\bibitem[{Holzwarth \& Jardine(2007)}]{Holzwarth07}
Holzwarth, V. \& Jardine, M. 2007, Astron. Astrophys., 463, 11

\bibitem[{Hubbard(1984)}]{Hubbard84}
Hubbard, W.~B. 1984, Planetary interiors (New York: Van Nostrand Reinhold Co.)

\bibitem[{Jiang {et~al.}(2003)Jiang, Ip, \& Yeh}]{Jiang03}
Jiang, I.-G., Ip, W.-H., \& Yeh, L.-C. 2003, Astrophys. J., 582, 449

\bibitem[{Khodachenko {et~al.}(2007{\natexlab{a}})Khodachenko, Lammer,
  Lichtenegger, Langmayr, Erkaev, Grie{\ss}meier, Leitner, Penz, Biernat,
  Motschmann, \& Rucker}]{Khodachenko06PSS}
Khodachenko, M.~L., Lammer, H., Lichtenegger, H. I.~M., {et~al.}
  2007{\natexlab{a}}, Planet. Space Sci., 55, 631

\bibitem[{Khodachenko {et~al.}(2007{\natexlab{b}})Khodachenko, Ribas, Lammer,
  Grie{\ss}meier, Leitner, Selsis, Eiroa, Hanslmeier, Biernat, Farrugia, \&
  Rucker}]{Khodachenko05}
Khodachenko, M.~L., Ribas, I., Lammer, H., {et~al.} 2007{\natexlab{b}},
  Astrobiology, 7, 167

\bibitem[{Laskar \& Correia(2004)}]{Laskar04}
Laskar, J. \& Correia, A. C.~M. 2004, in ASP Conference Series, Vol. 321,
  Extrasolar Planets: {T}oday and {T}omorrow, ed. J.-P. Beaulieu,
  A.~{Lecavelier des Etangs}, \& C.~Terquem, 401--410

\bibitem[{Lazio {et~al.}(2004)Lazio, Farrell, Dietrick, Greenless, Hogan,
  Jones, \& Hennig}]{Lazio04}
Lazio, T. J.~W., Farrell, W.~M., Dietrick, J., {et~al.} 2004, Astrophys. J.,
  612, 511

\bibitem[{{Le Qu\'{e}au} {et~al.}(1985){Le Qu\'{e}au}, Pellat, \&
  Roux}]{LeQueau85}
{Le Qu\'{e}au}, D., Pellat, R., \& Roux, A. 1985, Ann. Geophys., 3, 273

\bibitem[{Levrard {et~al.}(2007)Levrard, Correia, Chabrier, Baraffe, Selsis, \&
  Laskar}]{Levrard07}
Levrard, B., Correia, A. C.~M., Chabrier, G., {et~al.} 2007, Astron.
  Astrophys., 462, L5

\bibitem[{Lynden-Bell \& O'Dwyer(2001)}]{LyndenBell01preprint}
Lynden-Bell, D. \& O'Dwyer, J.~P. 2001, astro-ph/0104450

\bibitem[{Lynden-Bell \& Tout(2001)}]{LyndenBell01}
Lynden-Bell, D. \& Tout, C.~A. 2001, Astrophys. J., 558, 1

\bibitem[{MacDonald(1964)}]{MacDonald64}
MacDonald, G. J.~F. 1964, Rev. Geophys., 2, 467

\bibitem[{Majid {et~al.}(2006)Majid, Winterhalter, Chandra, Kuiper, Lazio,
  Naudet, \& Zarka}]{Majid05}
Majid, W., Winterhalter, D., Chandra, I., {et~al.} 2006, in Planetary Radio
  Emissions VI, ed. H.~O. Rucker, W.~S. Kurth, \& G.~Mann (Austrian Academy of
  Sciences Press, Vienna), 589--594

\bibitem[{Mann {et~al.}(1999)Mann, Jansen, MacDowall, Kaiser, \&
  Stone}]{Mann99}
Mann, G., Jansen, F., MacDowall, R.~J., Kaiser, M.~L., \& Stone, R.~G. 1999,
  Astron. Astrophys., 348, 614

\bibitem[{Marcy {et~al.}(1997)Marcy, Butler, Williams, Bildsten, Graham, Ghez,
  \& Jernigan}]{Marcy97}
Marcy, G.~W., Butler, R.~P., Williams, E., {et~al.} 1997, Astrophys. J., 481,
  926

\bibitem[{Mariani \& Neubauer(1990)}]{Mariani90}
Mariani, F. \& Neubauer, F.~M. 1990, in Physics of the Inner Heliosphere, ed.
  R.~Schwenn \& E.~Marsch, Vol.~1 (Berlin: Springer-Verlag), 183--206

\bibitem[{Murray \& Dermott(1999)}]{Murray99}
Murray, C.~D. \& Dermott, S.~F. 1999, Solar System Dynamics (Cambridge:
  Cambridge University Press)

\bibitem[{Nellis(2000)}]{Nellis00}
Nellis, W.~J. 2000, Planet. Space Sci., 48, 671

\bibitem[{Newkirk(1980)}]{Newkirk80}
Newkirk, Jr., G. 1980, in The Ancient Sun: Fossil Record in the Earth, Moon and
  Meteorites, ed. R.~O. Pepin, J.~A. Eddy, \& R.~B. Merrill, 293--320

\bibitem[{Parker(1958)}]{Parker58}
Parker, E.~N. 1958, Astrophys. J., 128, 664

\bibitem[{P\"{a}tzold \& Rauer(2002)}]{Paetzold02}
P\"{a}tzold, M. \& Rauer, H. 2002, Astrophys. J., 568, L117

\bibitem[{Peale(1999)}]{Peale99}
Peale, S.~J. 1999, Ann. Rev. Astron. Astrophys., 37, 533

\bibitem[{Preusse(2006)}]{PreussePHD05}
Preusse, S. 2006, PhD thesis, Technische Universit\"{a}t Braunschweig, {I}SBN
  3-936586-48-9, {C}opernicus-GmbH Katlenburg-Lindau

\bibitem[{Preusse {et~al.}(2005)Preusse, Kopp, B\"{u}chner, \&
  Motschmann}]{Preusse05}
Preusse, S., Kopp, A., B\"{u}chner, J., \& Motschmann, U. 2005, Astron.
  Astrophys., 434, 1191

\bibitem[{Preusse {et~al.}(2006)Preusse, Kopp, B\"{u}chner, \&
  Motschmann}]{Preusse06}
Preusse, S., Kopp, A., B\"{u}chner, J., \& Motschmann, U. 2006, Astron.
  Astrophys., 460, 317

\bibitem[{Pr\"{o}lss(2004)}]{Proelss04}
Pr\"{o}lss, G.~W. 2004, Physics of the Earth's Space Environment (Berlin:
  Springer-Verlag)

\bibitem[{Ryabov {et~al.}(2004)Ryabov, Zarka, \& Ryabov}]{Ryabov04}
Ryabov, V.~B., Zarka, P., \& Ryabov, B.~P. 2004, Planet. Space Sci., 52, 1479

\bibitem[{Saffe {et~al.}(2005)Saffe, G\'{o}mez, \& Chavero}]{Saffe05}
Saffe, C., G\'{o}mez, M., \& Chavero, C. 2005, Astron. Astrophys., 443, 609

\bibitem[{S\'{a}nchez-Lavega(2004)}]{Sanchez04}
S\'{a}nchez-Lavega, A. 2004, Astrophys. J., 609, L87

\bibitem[{Sato {et~al.}(2005)Sato, Fischer, Henry, Laughlin, Butler, Marcy,
  Vogt, Bodenheimer, Ida, Toyota, Wolf, Valenti, Boyd, Johnson, Wright, Ammons,
  Robinson, Strader, McCarthy, Tah, \& Minniti}]{Sato05}
Sato, B., Fischer, D.~A., Henry, G.~W., {et~al.} 2005, Astrophys. J., 633, 465

\bibitem[{Shkolnik {et~al.}(2003)Shkolnik, Walker, \& Bohlender}]{Shkolnik03}
Shkolnik, E., Walker, G. A.~H., \& Bohlender, D.~A. 2003, Astrophys. J., 597,
  1092

\bibitem[{Shkolnik {et~al.}(2004)Shkolnik, Walker, \& Bohlender}]{Shkolnik04}
Shkolnik, E., Walker, G. A.~H., \& Bohlender, D.~A. 2004, Astrophys. J., 609,
  1197

\bibitem[{Shkolnik {et~al.}(2005)Shkolnik, Walker, Bohlender, Gu, \&
  K\"{u}rster}]{Shkolnik05}
Shkolnik, E., Walker, G. A.~H., Bohlender, D.~A., Gu, P.-G., \& K\"{u}rster, M.
  2005, Astrophys. J., 622, 1075

\bibitem[{Showman \& Guillot(2002)}]{Showman02}
Showman, A.~P. \& Guillot, T. 2002, Astron. Astrophys., 385, 166

\bibitem[{Stevens(2005)}]{Stevens05}
Stevens, I.~R. 2005, Mon. Not. R. Astron. Soc., 356, 1053

\bibitem[{Tout {et~al.}(1996)Tout, Pols, Eggleton, \& Han}]{Tout96}
Tout, C.~A., Pols, O.~R., Eggleton, P.~P., \& Han, Z. 1996, Mon. Not. R.
  Astron. Soc., 281, 257

\bibitem[{Treumann(2000)}]{Treumann00}
Treumann, R.~A. 2000, in Geophysical monograph series, Vol. 119, Radio
  Astronomy at Long Wavelengths, ed. R.~Stone, K.~W. Weiler, M.~L. Goldstein,
  \& J.-L. Bougeret, 13--26

\bibitem[{Udry {et~al.}(2003)Udry, Mayor, \& Santos}]{Udry03}
Udry, S., Mayor, M., \& Santos, N.~C. 2003, Astron. Astrophys., 407, 369

\bibitem[{Voigt(1995)}]{Voigt95}
Voigt, G.-H. 1995, in Handbook of atmospheric electrodynamics, ed. H.~Volland,
  Vol.~II (CRC Press), 333--388

\bibitem[{Weber \& Davis(1967)}]{Weber67}
Weber, E.~J. \& Davis, Jr., L. 1967, Astrophys. J., 148, 217

\bibitem[{Winterhalter {et~al.}(2006)Winterhalter, Kuiper, Majid, Chandra,
  Lazio, Zarka, Naudet, Bryden, Gonzales, \& Treumann}]{Winterhalter06}
Winterhalter, D., Kuiper, T., Majid, W., {et~al.} 2006, in Planetary Radio
  Emissions VI, ed. H.~O. Rucker, W.~S. Kurth, \& G.~Mann (Austrian Academy of
  Sciences Press, Vienna), 595--602

\bibitem[{Wood(2004)}]{Wood04}
Wood, B.~E. 2004, Living Rev. Solar Phys., 1, 2, {U}RL:
  http://www.livingreviews.org/lrsp-2004-2, accessed on 23 February 2007

\bibitem[{Wood {et~al.}(2002)Wood, M\"{u}ller, Zank, \& Linsky}]{Wood02}
Wood, B.~E., M\"{u}ller, H.-R., Zank, G.~P., \& Linsky, J.~L. 2002, Astrophys.
  J., 574, 412

\bibitem[{Wood {et~al.}(2005)Wood, M\"{u}ller, Zank, Linsky, \&
  Redfield}]{Wood05}
Wood, B.~E., M\"{u}ller, H.-R., Zank, G.~P., Linsky, J.~L., \& Redfield, S.
  2005, Astrophys. J., 628, L143

\bibitem[{Yantis {et~al.}(1977)Yantis, Sullivan, \& Erickson}]{Yantis77}
Yantis, W.~F., Sullivan, III., W.~T., \& Erickson, W.~C. 1977, Bull. Am.
  Astron. Soc., 9, 453

\bibitem[{Zarka(2004)}]{Zarka04a}
Zarka, P. 2004, in ASP Conference Series, Vol. 321, Extrasolar planets: {T}oday
  and {T}omorrow, ed. J.-P. Beaulieu, A.~{Lecavelier des Etangs}, \&
  C.~Terquem, 160--169

\bibitem[{Zarka(2006)}]{Zarka06PREVI}
Zarka, P. 2006, in Planetary Radio Emissions VI, ed. H.~O. Rucker, W.~S. Kurth,
  \& G.~Mann (Austrian Academy of Sciences Press, Vienna), 543--569

\bibitem[{Zarka(2007)}]{Zarka06PSS}
Zarka, P. 2007, Planet. Space Sci., 55, 598

\bibitem[{Zarka {et~al.}(2004)Zarka, Cecconi, \& Kurth}]{Zarka04}
Zarka, P., Cecconi, B., \& Kurth, W.~S. 2004, J. Geophys. Res., 109, A09S15

\bibitem[{Zarka {et~al.}(2001{\natexlab{a}})Zarka, Queinnec, \&
  Crary}]{Zarka01cutoff}
Zarka, P., Queinnec, J., \& Crary, F.~J. 2001{\natexlab{a}}, Planet. Space
  Sci., 49, 1137

\bibitem[{Zarka {et~al.}(1997)Zarka, Queinnec, Ryabov, Ryabov, Shevchenko,
  Arkhipov, Rucker, Denis, Gerbault, Dierich, \& Rosolen}]{Zarka97}
Zarka, P., Queinnec, J., Ryabov, B.~P., {et~al.} 1997, in Planetary Radio
  Emissions IV, ed. H.~O. Rucker, S.~J. Bauer, \& A.~Lecacheux (Austrian
  Academy of Sciences Press, Vienna), 101--127

\bibitem[{Zarka {et~al.}(2001{\natexlab{b}})Zarka, Treumann, Ryabov, \&
  Ryabov}]{Zarka01}
Zarka, P., Treumann, R.~A., Ryabov, B.~P., \& Ryabov, V.~B. 2001{\natexlab{b}},
  Astrophys. Space Sci., 277, 293

\end{thebibliography}

\end{document}